\documentclass[10pt,prb,superscriptaddress,amsmath,amssymb, twocolumn]{revtex4-1}

\usepackage{amsmath}
\usepackage{mathrsfs}
\usepackage{graphicx}
\usepackage{dcolumn}
\usepackage{bm}
\usepackage{hyperref}

\renewcommand{\vec}{\mathbf}
\newcommand*\rfrac[2]{{}^{#1}\!/_{#2}}
\DeclareRobustCommand{\tr}{\genfrac{}{}{0pt}{}}

\begin{document}


\title{Damping of the Anderson-Bogolyubov mode in Fermi mixtures by spin and mass imbalance}

\author{Piotr Zdybel}
\author{Pawel Jakubczyk}
\affiliation{
 Institute of Theoretical Physics, Faculty of Physics, University of Warsaw\\
 Pasteura 5, 02-093 Warsaw, Poland
}%

\date{\today}

\begin{abstract}
We study the temporally nonlocal contributions to the gradient expansion of the pair fluctuation propagator for spin- and mass-imbalanced Fermi mixtures. These terms are related to damping processes of sound-like (Anderson-Bogolyubov) collective modes and are relevant for the structure of the complex pole of the pair fluctuation propagator.  We derive conditions under which damping occurs even at zero temperature for large enough mismatch of the Fermi surfaces. We compare our analytical results with numerically computed damping rates of the Anderson-Bogolyubov mode.
\end{abstract}

\maketitle


\section{Introduction} 

Progress in manipulating ultracold atomic setups\cite{bloch_2008, giorgini_2008, murkherjee_17, hueck_2018} opens opportunities to experimentally study many-body problems hardly accessible in conventional solid-state systems. In particular, the ability to engineer fermionic mixtures with different particle masses\cite{taglieber_2008, wille_2008, voigt_2009, tiecke_2010} and populations\cite{zwierlein_2006, partridge_2006, ketterle_2009,ong_2015, mitra_2016} motivates detailed studies of the influence of imbalance on superfluidity. For instance, exotic superfluid phases such as the interior gap (Sarma-Liu-Wilczek) superfluids\cite{sarma_1963, liu_2003} or the nonuniform Fulde-Ferrell-Larkin-Ovchinnikov (FFLO) states\cite{fulde_1964, larkin_1965} are theoretically possible to realize in two-component mixtures with different spin populations and masses of particles. Of substantial interest is also the imbalance-induced phase transition between the uniform superfluid and normal phases. By manipulating the mass imbalance, the tricritical point may be shifted\cite{parish_2007, baarsma_2010, radzihovsky_2010} or even expelled from the phase diagram giving rise to a stable quantum critical point (QCP).\cite{strack_2014, zdybel_2018} Another aspect concerns the possible quantum phase transition to the FFLO state.\cite{piazza_2016, pimenov_2018}\\
\indent In addition to thermodynamic properties of high interest are excitation spectra of such superfluid systems and their evolution upon increasing imbalance. Generally one expects the occurrence of a gapless (Anderson-Bogolyubov) sound-like branch (discussed in the present study) as well as a gapped amplitude mode. Note however that additional interesting features arise for example in two-band superfluids,\cite{iskin_2005, iskin_2007a, klimin_2015, klimin_2019a} or systems involving spin-orbit coupling.\cite{iskin_2011,liao_2012, seo_2012, zhang_2013, iskin_2019}\\
\indent One particularly interesting and generic problem concerns properties of Anderson-Bogolyubov (AB) modes,\cite{anderson_1958, bogoliubov_1958} also known as Nambu-Goldstone modes\cite{nambu_1960, goldstone_1961} in the low-momentum limit. According to the Goldstone theorem\cite{goldstone_1962} spontaneous breaking of continuous $U(1)$ symmetry for the Fermi gas results in low-energy sound-like collective excitations. These modes have been successfully observed in several experiments\cite{bartenstein_2004, altmeyer_2007, tey_2013, sidorenkov_2013, hoinka_2017}  and studied extensively in numerous theoretical papers\cite{engelbrecht_1997, marini_1998, ohashi_2003, combescot_2006, iskin_2007, hu_2007, diener_2008, klimin_2011, kurkjian_2016, tempere_2019, klimin_2019} in the last 20 years. However, most of these works pay relatively little attention to the spin- and mass-imbalance influence on the excitation spectra and their damping in particular. The dominant mechanism of damping in such systems is related to inelastic scattering of the Goldstone phonon from thermally excited fermionic quasiparticles.\cite{zou_2018} The damping rate vanishes in the limit $T\to0$\cite{zou_2018, shen_2015} because of the disappearance of the thermal cloud of quasiparticles.\cite{zhang_2011} This picture is consistent with the detailed analysis performed by Kurkjian and Tempere,\cite{kurkjian_2017} which shows that the process of absorption and emission of the AB phonon by fermionic quasiparticles leads to exponentially suppressed damping at low temperatures in presence of a gap. This temperature dependence is an essential characteristic of the so-called Landau damping\cite{bruus_2016} for gapped modes. However,\cite{matera_2017} large enough spin-polarization of the Fermi gas leads to enhancement of the damping factor even for relatively low temperatures. This suggests a relation between a mismatch of Fermi surfaces corresponding to the two particle species forming the mixture and the mechanism of the damping process.\\
\indent It is therefore worth taking a closer look at the problem of the impact of  spin- and mass-imbalance on the Landau damping. The present work contributes to an analytical understanding of the damping process by considering the structure of the Gaussian pair fluctuation (GPF) propagator in the low-momentum limit ($q\to 0$). We derive an inequality involving parameters of the system, giving a necessary condition to obtain a nonzero damping rate of the AB mode in a uniform s-wave superfluid in presence of both spin- and mass- imbalance.  Our central result  indicates that for large enough mismatch of the Fermi surfaces, the AB modes are damped even at $T=0$. We formulate an intuitive interpretation of this result and relate it to the mechanism of Landau damping.  We subsequently compare the conclusions drawn from the analytical results with numerically obtained complex poles ($z_q=\omega_q-i\Gamma_q/2$) of the GPF propagator, where the dispersion relation of the collective mode and its damping rate are given, respectively, by the real and imaginary part of $z_q$.\\
\indent The paper is organized as follows. In Sec.~II, we  introduce the considered model using the path-integral formalism and discuss the structure of the GPF~propagator. In Sec.~III, we employ the gradient expansion to extract the leading terms of the GPF~propagator responsible for damping. We formulate conditions under which Landau damping is active even for $T=0$ and present an intuitive interpretation of this result.
Sec.~IV contains a numerical study of the poles of the GPF propagator. We discuss the obtained dispersion relations and damping rates of the AB modes for different realizations of spin- and mass-imbalance and compare the results with analytical expressions from Sec.~III. We summarize the paper and give a perspective for future studies in Sec.~V.

\section{Pair fluctuation propagator}

We consider a two-component spin-polarized Fermi mixture with unequal masses in thermodynamic equilibrium. Particles with opposite spins interact  via an attractive contact potential $\mathcal{V}(\vec{x},\vec{y})=g\delta(\vec{x}-\vec{y})$, where $g<0$. Utilizing the path integral formalism,\cite{altland_2010} we obtain the grand canonical partition function represented as a functional integral over the Grassmann fields $\{\bar{\psi}^\sigma_x, \psi^\sigma_x \}$
 \begin{equation}
Z=\int\mathcal{D}[\bar{\psi}^\sigma_x, \psi^\sigma_x]\exp(-S_\psi)\;, 
\end{equation} 
where 
\begin{equation}
S_\psi=\int_x\left\{\sum_{\sigma} \bar{\psi}^\sigma_x\left[\partial_\tau+\hat{\mathcal{K}}^\sigma\right]\psi^\sigma_x+g\bar{\psi}^+_x\bar{\psi}^-_x\psi^-_x\psi^+_x\right\}
\label{action}
\end{equation}
is the fermionic action. Throughout the paper we put $\hslash=1$ and $k_B=1$. In the above equations, we use the following notation: $x=(\tau, \mathbf{x})$, $\int_x(\cdot)=\int_0^\beta\mathrm{d}\tau\int\mathrm{d}^d \vec{x}\,(\cdot)$, $\beta=1/T$ and $\hat{\mathcal{K}}^\sigma=-\nabla^2_\vec{x}/2m_\sigma-\mu_\sigma$, where $m_\sigma$ and $\mu_\sigma$ are the mass and chemical potential of a particle with spin $\sigma\in\{+, -\}$, respectively. The presence of two distinct Fermi surfaces in this problem makes it convenient to introduce the imbalance parameters. We define $\zeta=\frac{r-1}{r+1}$, where $r=m_-/m_+$. We also use $h=(\mu_+-\mu_-)/2$ as the 'Zeeman' field, which measures the spin polarization and $\mu=(\mu_++\mu_-)/2$ as the average chemical potential.\\
\indent The standard procedure to analyze the Gaussian fluctuations is to integrate out the Grassmann fields $\psi^\sigma_x$ by introducing an auxiliary field $\eta_x$ via the Hubbard-Stratonovich transformation,\cite{engelbrecht_1997, iskin_2007} which decouples the interaction term in Eq. (\ref{action}) into the Cooper channel.\cite{altland_2010} Afterwards, we expand $\eta_q$ (in reciprocal space)  around the mean-field value of the superfluid gap $\Delta$,  in such a way that $\eta_q=\sqrt{\beta V}\,\Delta\,\delta_{q,0}+\phi_q$, where $V$ is the volume of the system and $\phi_q$ describes fluctuations of the order parameter. Thus, the expansion to the quadratic order in $\phi_q$ leads to a Gaussian action. For a comprehensive discussion of this procedure we refer to the paper by Iskin and S\'a de Melo.\cite{iskin_2007} Alternatively, the same result can be obtained using diagrammatic theory by a resummation of the infinite subclass of ladder diagrams (the random phase approximation).\cite{nozieres_1985, pieri_2004}\\
\indent As a result, we obtain the partition function within the GPF approximation, which  is given by
\begin{equation}
Z=Z_{MF}\int\mathcal{D}[\phi]\exp\left(-\frac{\beta V}{2}\int_q\Phi^*_q\mathbb{F}^{-1}_q\Phi_q\right),
\label{Z}
\end{equation}
where $Z_{MF}$ is the mean-field part of the partition function, $\Phi^*_q=[\phi_q^*, \phi_{-q}]$, $\Phi_q=[\phi_q, \phi^*_{-q}]^T$ and $\mathbb{F}_q$ is the GPF propagator matrix. In Eq. (\ref{Z}), $q=(iq_m, \vec{q})$ collects a bosonic Matsubara frequency [$q_m=2\pi m/\beta$ ($m\in\mathbb{Z}$)] and the ($d$-dimensional) wave vector $\vec{q}$. We also introduce the fermionic analogue $k=(ik_n, \vec{k})$, where $k_n=2\pi(n+\rfrac{1}{2})/\beta$ ($n\in\mathbb{Z}$). We use the shorthand notation: $\int_q(\cdot)=\frac{1}{\beta}\sum_{q_m}\int\frac{\mathrm{d}^d \vec{q}}{(2\pi)^d}\,(\cdot)$. Matrix elements of the inverse GPF propagator are expressed by normal ($\mathscr{G}^\sigma_k$) and anomalous ($\mathscr{F}_k$) components of the Green function matrix:\cite{abrikosov_2016}
\begin{align}
[\mathbb{F}^{-1}_q]_{1,1}&\equiv M_{1,1}(q)=\frac{1}{g}-\int_k \mathscr{G}^{+}_{k+q}\mathscr{G}^{-}_{-k}\;, \label{M11}\\
[\mathbb{F}^{-1}_q]_{1,2}&\equiv M_{1,2}(q)=\int_k \mathscr{F}_{k+q}\mathscr{F}_{-k}\;, \label{M12}
\end{align}
where $M_{1,1}(q)=M_{2,2}(-q)$, $M_{1,2}(q)=M^*_{2,1}(q)$, while the BCS-like Green functions are given by 
\begin{align}
\mathscr{G}^{+}_{k}&=\frac{u_k^2}{ik_n-E_k^+}+\frac{v_k^2}{ik_n-E_k^-}\;, \label{G+}\\
-\mathscr{G}^{-}_{-k}&=\frac{v_k^2}{ik_n-E_k^+}+\frac{u_k^2}{ik_n-E_k^-}\;, \label{G-}\\
\mathscr{F}_{k}&=u_k v_k^* \left(\frac{1}{ik_n-E_k^-}-\frac{1}{ik_n-E_k^+} \label{F}\right)\;.
\end{align}
Here we use the BCS coherence factors given by $u^2_k=(1+\xi_k/E_k)/2$ and $v_k^2=1-u_k^2$, where  $\xi_k=\vec{k}^2/2m-\mu$, $E_k=\sqrt{\xi_k^2+|\Delta|^2}$ and $m=\frac{2r}{r+1}m_+$. $E_k^\sigma$ is an excitation energy of quasi-particle branches in the superfluid phase:
\begin{align}
E_k^\sigma=\zeta\xi_k-(h-\zeta\mu)+\sigma E_k\;.
\label{spec}
\end{align}
It is worth noting that $h=\zeta\mu$ corresponds to a situation where the two Fermi surfaces coincide. Therefore, $h-\zeta\mu$ measures the mismatch of Fermi spheres due to spin- and mass-imbalance. For further discussion of the matrix elements of $\mathbb{F}^{-1}_q$, see appendix A.\\
\indent In order to obtain the mean-field value of the superfluid gap $\Delta$, we consider the contribution $\Omega_{MF}=-T\ln Z_{MF}$ to the grand-canonical potential. This is given by\cite{zdybel_2018}
\begin{equation}
\Omega_{MF}=V\min_\Delta\left\{-\frac{|\Delta|^2}{g}+T\int_\vec{k}\sum_\sigma \ln f(-E_k^\sigma)\right\},
\label{MF}
\end{equation}
where $\int_\vec{k}(\cdot)=\int\frac{\mathrm{d}^d \vec{k}}{(2\pi)^d}\,(\cdot)$ and $f(x)=1/(\exp(\beta x)+1)$. The order parameter minimizing the grand-canonical potential for a given set of parameters is identified as the expectation value of the superfluid gap $\Delta$.

\section{Damping of collective modes}

We now set out to analyze the complex pole of the GPF propagator. For this purpose, we expand the matrix elements of $\mathbb{F}_q^{-1}$ in the low-momentum limit and extract the relevant nonlocal terms, which are related to the damping process. An analogous strategy is applied to derive the Hertz-Millis-Moriya action in the context of quantum phase transitions in itinerant electron systems.\cite{hertz_1976, nagaosa_1998, lohneysen_2007, continentino_2017} In this case, the nonlocal term $|q_m|/\gamma_q$ appearing in the Gaussian action after the expansion for small $|\vec{q}|$ and $|q_m|/|\vec{q}|$ is related to the Landau damping of collective spin fluctuations by particle-hole excitations.\cite{lohneysen_2007} This term is responsible for the occurrence of the complex pole of the propagator of paramagnons and $\gamma_q$ is its damping rate.\cite{hertz_1976}  The noticeable structural resemblance between the Hertz approach and our problem encouraged us to exploit this procedure to investigate the Landau damping of the AB mode in the spin- and mass-imbalance Fermi mixture.\\
\indent We begin with a brief discussion of the gradient expansion along the line of Refs.~\onlinecite{diener_2008} and \onlinecite{klimin_2011}.
We obtain a low-momentum and low-frequency  expansion of the matrix elements $M_{j, l}(iq_m ; \vec{q})$ [see Eq. (\ref{M11A}) and (\ref{M12A})] up to the second-order in powers of $q_m$ and $\vec{q}$:
\begin{align}
M_{1,1}(iq_m; \vec{q})&=M_{1,1}(0; \vec{0})+A\vec{q}^2+iBq_m+C q^2_m\;, \label{M11E}\\
 M_{1,2}(iq_m; \vec{q})&=M_{1,2}(0; \vec{0})+D\vec{q}^2+Eq_m^2\;. \label{M12E}
\end{align}
The expressions for the coefficients of the gradient expansion are presented in appendix B. This procedure neglects terms proportional to $|q_m|/|\vec{q}|$, which are crucial in the description of damping. To identify them, we analyze the full expression for $M_{1,1}(iq_m; \vec{q})-M_{1,1}(0,\vec{q})$. Using the notation described in detail in appendix A, we start from the following form:
\begin{align}
M_{1,1}(iq_m; \vec{q})-M_{1,1}(0;\vec{q})=\;\;\;\;\;\;\;\;\;\;\nonumber\\-\int_\vec{k}\sum_{\sigma, \sigma'}\mathcal{C}_{k, q}^{\sigma, \sigma'}\frac{f_{k, q}^{\sigma, \sigma'}}{q^2_m+(E_{k, q}^{\sigma, \sigma'})^2}\left(iq_m-\frac{q_m^2}{E_{k, q}^{\sigma, \sigma'}}\right).
\label{M11S}
\end{align}
The leading contribution involving \emph{both} small frequency and momentum [and therefore not included in the expansion of Eq. (\ref{M11E})] comes from the second term in the bracket in Eq.~(\ref{M11S}) for the elements with $\sigma=\sigma'$ and is given by
\begin{align}
-\int_{\vec{k}}u_k^2v_k^2\sum_\sigma f'(E_k^\sigma)\frac{q_m^2}{a^2_\sigma(|\vec{k}|)\cos^2\theta |\vec{q}|^2+q_m^2}\;,
\label{EX}
\end{align}
where $\cos\theta=\frac{\vec{k}\cdot\vec{q}}{|\vec{k}||\vec{q}|}$, $f'(x)=-\frac{\beta}{4}\cosh^{-2}(\beta x/2)$ and $a_\sigma(|\vec{k}|)=(\frac{\zeta}{m}+\sigma\frac{\xi_k}{mE_k})|\vec{k}|$. All the other contributions (as long as $|\Delta|>0$) in the expansion of  Eq. (\ref{M11S}) are either of higher order or included in Eq. (\ref{M11E}).\\
\indent  In the next step we perform integration over the angular variable $\theta$, considering separately the cases  $d=2$ and $d=3$. We assume that the ratio $|q_m|/|\vec{q}|$ is small.\cite{hertz_1976, nagaosa_1998} Note that this is possible only if the mode in question is gapless.\cite{diener_2008} In Eq.~(\ref{EX}) we make the change of variables, $|\vec{k}|\to\varepsilon=\vec{k}^2/2m\geq0$ and carry out the integration. This yields
\begin{align}
-\frac{|q_m|}{|\vec{q}|}\int\mathrm{d}\varepsilon \; c_d D_d(\varepsilon) u_\varepsilon^2v_\varepsilon^2\sum_\sigma\frac{f'(E_\varepsilon^\sigma)}{|a_\sigma(\varepsilon)|}=\frac{|q_m|}{\gamma_q}\;,
\label{gammaT}
\end{align}
where $c_d$ is equal $1$ for $d=2$ and $\pi/2$ for $d=3$. We also introduce the density of states per spin $D_d(\varepsilon)$, where $D_2(\varepsilon)=m/2\pi$ and $D_3(\varepsilon)=\frac{\sqrt{2m^3}}{2\pi^2}\varepsilon^{1/2}$. Eq.~(\ref{gammaT}) defines the quantity $\gamma_q$.  An analogous procedure performed for the matrix element $M_{1,2}(iq_m, \vec{q})$ results in the same expression as in Eq.~(\ref{gammaT}). We conclude that relations (\ref{M11E}) and (\ref{M12E}) should be supplemented by the nonlocal contributions obtained in Eq.~(\ref{gammaT}).\\
\begin{figure}[h]
\centering 
    \includegraphics[width=0.4 \textwidth]{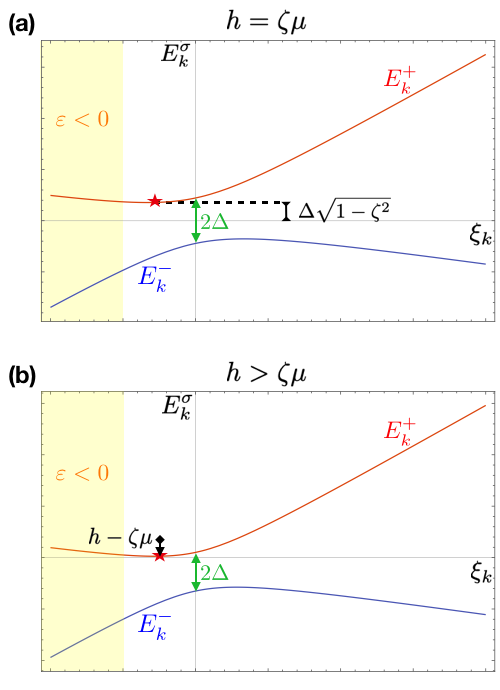} 
    \caption{Schematic illustration of the fermionic quasiparticle spectrum $E_k^\sigma$ [see Eq. (\ref{spec})] as a function of $\xi_k=\varepsilon-\mu$, when the minimum of $E_k^+$ is located in the physical region ($\varepsilon\geq 0$). The red star indicates the minimum of $E_k^+$  and  in the shaded area we have $\varepsilon<0$. (a) In this case, the two Fermi spheres coincide ($h-\zeta\mu=0$). The lower branch of the quasiparticle spectrum is fully occupied while the upper branch is empty. In this case, the Landau damping involves the inelastic scattering of the Goldstone phonon from thermally excited quasiparticles and becomes inactive for $T \to 0$. (b) The mismatch of the Fermi surfaces leads to nonzero occupation of the upper branch, whenever the condition $h-\zeta\mu\geq\Delta\sqrt{1-\zeta^2}$ is fulfilled. In this case, the damping process occurs also at zero temperature.}
    \label{pic1}
\end{figure}
\indent We now consider the form of $\gamma^{-1}_q$ in the zero-temperature limit. Using $f'(E_\varepsilon^\sigma)\to -\delta(-E_\varepsilon^\sigma)$ for $T\to0$, we obtain
\begin{align}
\gamma^{-1}_q=\frac{1}{|\vec{q}|}\int\mathrm{d}\varepsilon \; c_d D_d(\varepsilon) u_\varepsilon^2v_\varepsilon^2\sum_\sigma\frac{\delta(-E_\varepsilon^\sigma)}{|a_\sigma(\varepsilon)|}\;.
\label{gamma}
\end{align}
Taking advantage of the identity
\begin{align}
\delta[h(x)]=\sum_i \frac{\delta(x-x_i)}{|h'(x_i)|}\;,
\label{delta}
\end{align}
where $x_i$ are roots of $h(x)$, we further simplify Eq.~(\ref{gamma}). The equation $E_\varepsilon^\sigma=0$ has two solutions, which are identical for $\sigma=+$ and $\sigma=-$ and given by
\begin{align}
\varepsilon_{1,2}=\frac{\mu-\zeta h \pm\sqrt{(h-\zeta\mu)^2-\Delta^2(1-\zeta^2)}}{1-\zeta^2}\;.
\label{root}
\end{align}
Since $\varepsilon=\vec{k}^2/2m$ is non-negative, we pick only roots fulfilling $\varepsilon_i\geq0$. Making use of Eq.~(\ref{delta}), we integrate over $\varepsilon$, which leads to 
\begin{align}
\gamma^{-1}_q=\frac{1}{|\vec{q}|} \sum_{\tr{i=1,2;}{\mathrm{if}\,\varepsilon_i\geq 0}} c_d D_d(\varepsilon_i) u_{\ell_{i}}^2 v_{\ell_{i}}^2\sum_\sigma L_\sigma^{-1}(\varepsilon_i)\;,
\label{gamT0}
\end{align}
where $\ell_i=\sqrt{2 m\varepsilon_i}$ and $L_\sigma(\varepsilon)=\sqrt{m/2\varepsilon}|a_\sigma(\ell)|^2$. \\ 
\indent We now discuss implications of Eq.~(\ref{gamT0}). First of all, we observe that $\gamma^{-1}_q\sim u_k^2 v_k^2=\Delta^2/4E_k^2$. Therefore, $\gamma^{-1}_q$ vanishes in the limit $\Delta \to 0$. Moreover, $ u_k^2 v_k^2$ takes maximal value for $k=\sqrt{2m\mu}$.\\
\begin{figure}[h]
\centering 
    \includegraphics[width=0.4 \textwidth]{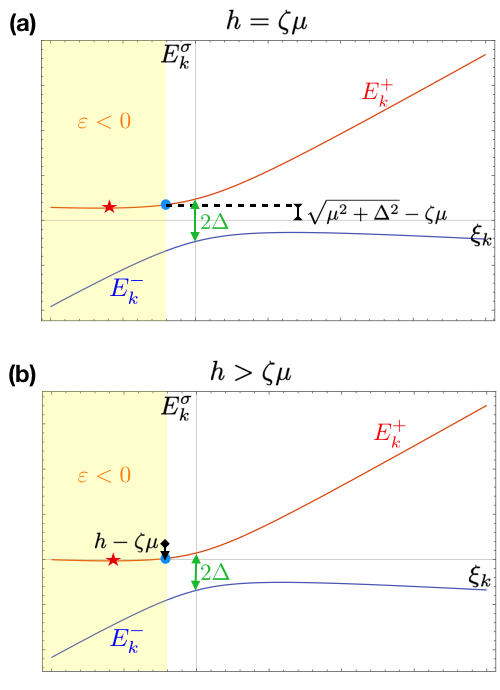} 
    \caption{The fermionic quasiparticle spectrum $E_k^\sigma$ [see Eq. (\ref{spec})] as a function of $\xi_k=\varepsilon-\mu$, when the minimum of $E_k^+$ is located in the unphysical region ($\varepsilon< 0$). The red star indicates the minimum of $E_k^+$, the blue dot corresponds to $\varepsilon=0$, and  in the shaded area we have $\varepsilon<0$. (a) In this case, the Fermi spheres coincide ($h-\zeta\mu=0$). The upper branch of the quasiparticle spectrum is empty (the lower one is fully occupied), therefore the Landau damping is possible only due to the presence of thermal excitations from $E_\varepsilon^-$ to $E_\varepsilon^+$. (b) The mismatch of the Fermi surfaces leads to the nonzero occupation of the upper branch, when the condition $h-\zeta\mu\geq\sqrt{\mu^2+\Delta^2}-\zeta\mu$ is met. In this case, Landau damping is active even at $T=0$.}
    \label{pic2}
\end{figure}
\indent As we mentioned above, $\varepsilon_i$ in Eq. (\ref{gamT0}) should be non-negative. This leads to the necessary condition for the occurrence of Landau damping. Indeed, when the requirement
\begin{align}
|h-\zeta\mu|>\Delta\sqrt{1-\zeta^2}
\label{con1}
\end{align}
is met, there are two real roots of $E_\varepsilon^\sigma=0$. In particular, whenever the two Fermi spheres coincide ($h=\zeta\mu$) we see that $\gamma_q^{-1}=0$ for $T=0$ and the Goldstone mode is not damped. Nonetheless, compliance with the condition in Eq. (\ref{con1}) does not guarantee fulfillment of $\varepsilon_i\geq 0$. For simplicity, let us now focus on the situation, where $h-\zeta \mu\geq 0$ and then consider $\varepsilon\geq 0$. This leads to
\begin{equation}
\begin{cases}
h-\zeta\mu\geq\Delta\sqrt{1-\zeta^2} \;\;\;\;\mathrm{for }\;\mu\geq\frac{\zeta\Delta}{\sqrt{1-\zeta^2}}\;, \\
h\geq\sqrt{\mu^2+\Delta^2} \;\;\;\;\;\;\;\;\;\;\;\;\mathrm{for }\;\mu<\frac{\zeta\Delta}{\sqrt{1-\zeta^2}}\;.
\end{cases} 
\label{con2}
\end{equation}
The first inequality, in the above condition, assures the existence of at least one positive zero of $E_\varepsilon^+$ (see Fiq. \ref{pic1}) and the second one corresponds to exactly one zero (see Fig.~\ref{pic2}).\\
\indent We can now interpret the obtained results in the context of the mechanism of  Landau damping. Let us for now  assume that $r\in [1,\infty[$ so that $\zeta\in[0,1[$. The quasiparticle spectrum has two branches [see Eq. (\ref{spec})]. If the two Fermi spheres coincide ($h-\zeta\mu=0$), then the lower branch $E_\varepsilon^-$ is filled, and the upper one $E_\varepsilon^+$ is empty.\cite{zhang_2011} In this case, the Landau damping is present only at nonzero temperatures (and is exponentially suppressed). The Goldstone phonon inelastically scatters thermally excited quasiparticles in the upper branch of the spectrum. Cranking up the mismatch of the Fermi surfaces leads to nonzero occupancy of fermions on $E_\varepsilon^+$  even at $T=0$. Therefore, the Landau damping is present also at $T=0$. We depicted this situation in Figs.~\ref{pic1}~and~\ref{pic2}. The position of the minimum of $E_\varepsilon^+$ is given by $\varepsilon_{min}=\mu-\zeta \Delta/\sqrt{1-\zeta^2}$ and at that point $E_{\varepsilon}^+$ is equal to $\Delta\sqrt{1-\zeta^2}$. Whenever $\varepsilon_{min}\geq 0$, the minimal mismatch of the Fermi spheres, which leads to nonzero occupancy of quasiparticles in the upper branch, is given by $\Delta\sqrt{1-\zeta^2}$ (see Fig.~\ref{pic1}). If $\varepsilon_{min}< 0$, the minimal mismatch leading to nonzero occupancy in $E_\varepsilon^+$  is given by a value of $E_\varepsilon^+$ for $\varepsilon=0$ (see Fig.~\ref{pic2}).This provides an interpretation of Eq. (\ref{con2}). We see that in the limit $r\to\infty$ the obtained condition is independent of $\zeta$ and is given by $h\geq \sqrt{\mu^2+\Delta^2}$, whereas for $r=1$ the considered condition is given by $h\geq\Delta$, which is consistent with the results of Ref.~\onlinecite{matera_2017}. We emphasize that the above results require the presence of a superfluid gap ($\Delta>0$). As we show in the next section for experimentally motivated choices of the mass imbalance parameter $r$, the inequality (\ref{con2}) becomes fulfilled for $h$ substantially lower than the critical value $h_c$, such that damping is present in a~broad region of the phase diagram within the superfluid phase. We also note that the occurrence of damping is not interrelated with the order of the phase transition to the normal phase.\\
\indent We close this section by considering the Landau damping in the proximity of a QCP,\cite{zdybel_2018, strack_2014} which can be generated for a wide range of system parameters. At mean-field level, in the limit $h\to h_c^-$ ($h_c$ is the critical value of~$h$), the superfluid gap $\Delta$ goes continuously to $0$. In the vicinity of the QCP, the condition $\varepsilon_1\geq 0$ yields
\begin{equation}
h+\mu-\frac{1+\zeta}{2(h-\zeta\mu)}\Delta^2+\mathcal{O}(\Delta^4) \geq 0.
\end{equation}
According to Ref. \onlinecite{zdybel_2018}, the above condition is always fulfilled for $r>3.01$ and $h>\mu\geq0$. Therefore, in situations where the quantum phase transition is continuous, the Landau damping is unavoidably present in the proximity of the QCP.

\section{Numerical results}

In this section we numerically study the damping of the collective mode by analyzing the complex pole of the GPF propagator. This amounts to finding complex roots of the following equation:\cite{engelbrecht_1997, tempere_2019, klimin_2019,  matera_2017}
\begin{equation}
\det\mathbb{F}^{-1}(iq_m\mapsto z_q; \vec{q})=0,
\label{det}
\end{equation}
where $z_q=\omega_q-i\Gamma_q/2$, $\omega_q$ is the dispersion relation and $\Gamma_q$ is the damping rate. The matrix elements $M_{j,l}(z_q, \vec{q})$ of $\mathbb{F}^{-1}_q$ have a branch cut along the real axis.\cite{klimin_2019}  We should perform the analytic continuation of $M_{j,l}(z_q, \vec{q})$ from the upper to the lower complex half-plane, which results in a~transition to another Riemann sheet. We proceed along the way described by Nozi\`eres.\cite{nozieres_2018} We consider the quantity $A_{j,l}(\omega; \vec{q})$:
\begin{equation}
A_{j,l}(\omega; \vec{q})=-\frac{1}{\pi}\mathrm{Im}\,M_{j,l}^{(R)}(\omega; \vec{q}),
\label{sf}
\end{equation}
where the index $(R)$ means that we take the retarded matrix element $M_{j,l}(iq_m\mapsto\omega+i0^+;\vec{q})$.  Then the matrix elements $\tilde{M}_{j,l}$ analytically continued to the lower half-plane are given by\cite{klimin_2019, matera_2017}
\begin{equation}
\tilde{M}_{j,l}^{(A)}(\omega; \vec{q})=M_{j,l}^{(A)}(\omega; \vec{q})-2\pi i A_{j,l}(\omega; \vec{q}),
\label{acm}
\end{equation}
where the index $(A)$ denotes the advanced counterpart of the matrix element ($iq_m\mapsto\omega-i0^+$). $\tilde{M}_{j,l}$ thus obtained can be extended in such a way that $\omega\mapsto z=\omega-i\Gamma/2$, where $\Gamma>0$.\\
\begin{figure}[h]
\centering 
    \includegraphics[width=0.45 \textwidth]{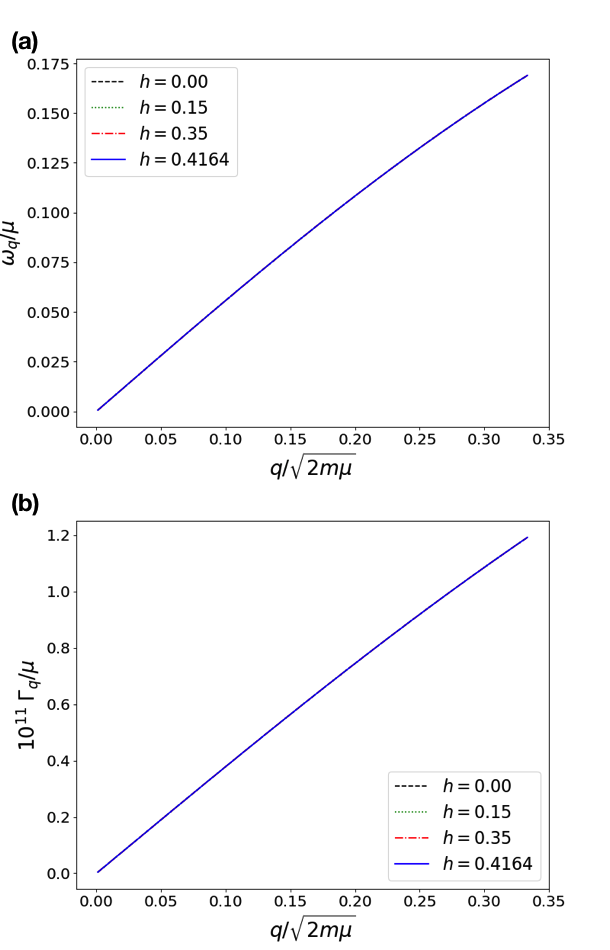} 
    \caption{(a) Dispersion relations $\omega_q$ of Goldstone modes (in units of $\mu$) as a function of momentum for $r=1$ and several values of $h$. For small enough values of $|\vec{q}|/\sqrt{2m\mu}$, we have $\omega_q=v_s|\vec{q}|+\mathcal{O}(|\vec{q}|^3)$. (b) Analogous figure for damping rates $\Gamma_q$ (in units of $\mu$). All the curves  coincide because of the weak dependence of the gap $\Delta$ on $h$. The damping factors are negligibly small. The plot parameters are $m_+=1$, $r=1$, $\mu=0.5$, $T=0.005$, $g=-2.0$, and $\Lambda=10$, where $\Lambda$ is the upper momentum cutoff.}
    \label{r1}
\end{figure}
\begin{figure}[h]
\centering 
    \includegraphics[width=0.45 \textwidth]{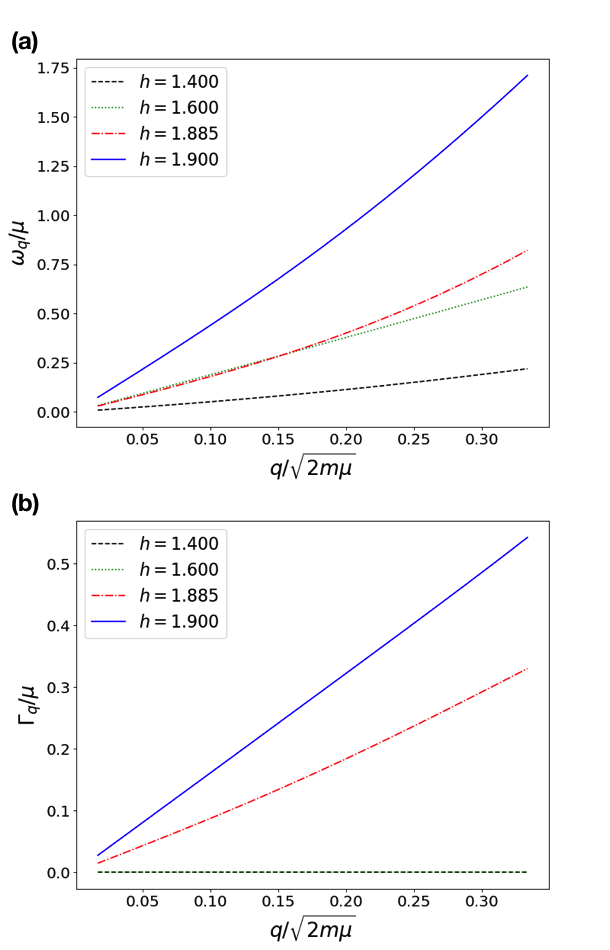} 
    \caption{(a) Dispersion relations $\omega_q$ of Goldstone modes (in units of $\mu$) as a function of momentum for $r=6.67$ and several values of $h$. For small enough values of $|\vec{q}|/\sqrt{2m\mu}$, we have that $\omega_q=v_s|\vec{q}|+\mathcal{O}(|\vec{q}|^3)$. (b) Analogous figure for damping rates $\Gamma_q$ (in units of $\mu$). Landau damping becomes active above $h\approx1.59$, which is substantially lower than the critical value $h_c\approx 1.97$. The plot parameters are $m_+=1$, $r=6.67$, $\mu=0.1$, $T=0.04$, $g=-1.4$ , and $\Lambda=10$.}
    \label{r6_67}
\end{figure}
\begin{figure}[h]
\centering 
    \includegraphics[width=0.45 \textwidth]{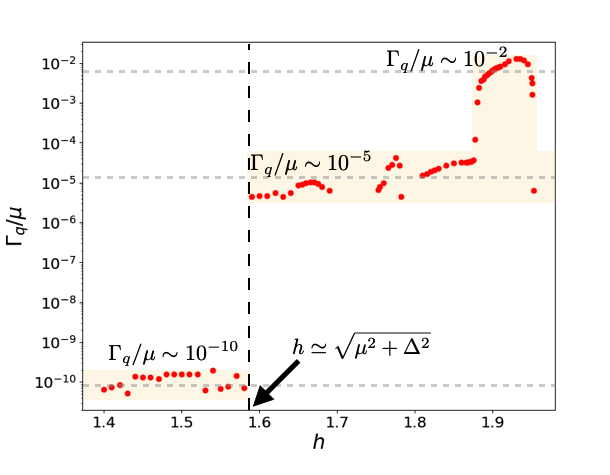} 
    \caption{Damping rates $\Gamma_q$ of Goldstone modes (in units of $\mu$ and with a logarithmic scale) as a function of the 'Zeeman' field for $r=6.67$ and $|\vec{q}|/\sqrt{2m\mu}=4.24\cdot10^{-3}$. Damping is active for $h\gtrsim\sqrt{\Delta^2+\mu^2}$. Moreover, for values of $h$ corresponding to the vicinity of the phase transition ($h\in[1.88, 1.95]$), damping becomes significantly stronger. The plot parameters are $m_+=1$, $r=6.67$, $\mu=0.1$, $T=0.04$, $g=-1.4$ , and $\Lambda=10$. A continuous phase transition between superfluid and normal phase occurs at $h_c=1.9706$.}
    \label{Pic3}
\end{figure}
\indent Using the procedure specified above (see Ref. \onlinecite{klimin_2019, matera_2017, nozieres_2018}) we discuss the numerically obtained dispersion relations $\omega_q$ and damping rates $\Gamma_q$ for $r=1.0$ and $r=6.67$, varying the 'Zeeman' field $h$ at $T\to 0$. We consider the~three-dimensional case ($d=3$). We \emph{a posteriori} check fulfillment of the condition $v_s=\lim_{|\vec{q}|\to0}\omega_q/|\vec{q}|<1$ (compare Sec.~III). This condition ensures that the assumptions made in derivation of Eq.~(\ref{gamma}) are justified. We begin with the mass-balanced case ($r=1$). The phase transition between the normal and superfluid phases is generically discontinuous for $r<3.01$ and $T\to 0$ at the mean-field level.\cite{zdybel_2018} Therefore, we expect that for $r=1$ the change of $\Delta$ as a function of $h$ should be modest up to the occurrence of the phase transition. In this case, the frequencies and damping factors of the Goldstone mode as a~function of momentum $|\vec{q}|$ are shown in Fig.~\ref{r1}. The results are not affected by varying $h$ between $0.0$ and $0.4164$, where the discontinuous phase transition takes place. The reason for this is negligible change of $\Delta$ upon approaching the transition point. For all values of $h$ considered in Fig.~\ref{r1} the ratio $T/\Delta<0.01$, which means that the thermal excitations should be negligible. For $r=1$ ($\zeta=0$) we always obtain $\mu>\zeta\Delta/\sqrt{1-\zeta^2}=0$. Therefore, the condition in  Eq.~(\ref{con2}) takes the form $h\geq\Delta$, which is never fulfilled for the discussed situation. That implies that the Landau damping is absent in the limit $T\to 0$ in compliance with the numerical results shown in Fig.~\ref{r1} (the numerically obtained damping rates, in this case, are of the order of $10^{-11}$). These results are consistent with Refs. \onlinecite{klimin_2011} and \onlinecite{tempere_2019}.\\
\indent We now examine the mass-imbalanced case, fixing the mass ratio $r=6.67$ corresponding to a $^6$Li and $^{40}$K mixture.\cite{iskin_2007} We choose the parameters so that the system hosts a QCP in its phase diagram. Note however that this is of no relevance for the occurrence of Landau damping. In this case, the QCP at the mean-field level is located at $h_c=1.9706$. The corresponding dispersion relations and damping rates of the Goldstone phonons are shown in Fig.~\ref{r6_67} for a few values of $h<h_c$. Since $T/\Delta<0.04$  for all $h$ in Fig.~\ref{r6_67}, we can reliably neglect thermal excitations. Furthermore, it turns out that for all the considered values of the 'Zeeman' field we can apply the criterion $h\geq\sqrt{\mu^2+\Delta^2}$ for the occurrence of the Landau damping [see Eq.~(\ref{con2})]. First we observe that for $h=1.4$ the above condition is not met. Thus, the upper branch of the quasiparticle spectrum is not populated, which means that the damping mechanism discussed in Sec. III is inactive. This is in agreement with the numerical results, which show that the damping rate is of the order of $10^{-10}$. Second, for $h=1.6$ we observe that $h\simeq\sqrt{\mu^2+\Delta^2}=1.59$. The obtained numerical values of $\Gamma_q/\mu$ are of the order of $10^{-5}$, which is way larger than the value obtained for $h=1.4$. In the remaining cases, the considered condition is fulfilled. In consequence, the upper branch of the excitation spectrum is partially occupied by quasiparticles even for $T\to 0$. Therefore, Goldstone modes can be absorbed by fermionic excitations and the Landau damping is present. As we see in Fig.~\ref{r6_67}, this prediction is consistent with numerical results. Moreover, the dependence $\Gamma_q/\mu$ on $h$ is shown in Fig.~\ref{Pic3} for $|\vec{q}|/\sqrt{2m\mu}=4.24\cdot10^{-3}$. We see that the activation of damping occurs precisely for the predicted value of $h$.\\
\begin{figure}[h]
\centering 
    \includegraphics[width=0.45 \textwidth]{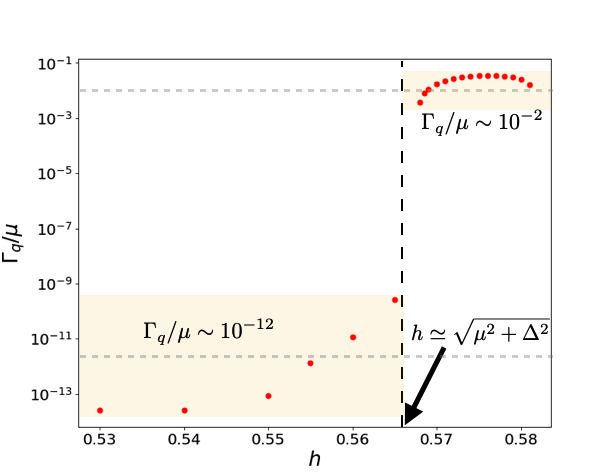} 
    \caption{Damping rates $\Gamma_q$ of Goldstone modes (in units of $\mu$ and with a logarithmic scale) as a function of the 'Zeeman' field for $r=3.47$ and $|\vec{q}|/\sqrt{2m\mu}=0.0179$. Damping is active for $h\gtrsim\sqrt{\Delta^2+\mu^2}$. In this case, phase diagram hosts a 1st~order phase transition for $h_{I}=0.59106$. The plot parameters are $m_+=1$, $r=3.47$, $\mu=0.1$, $T=0.002$, $g=-1.4$ , and $\Lambda=10$.}
    \label{Pic4}
\end{figure}
\indent The choice of parameters discussed above corresponds to a situation, where the phase transition between the normal and superfluid states is continuous. The obtained conclusion is however not sensitive to the order of the transition. To demonstrate this explicitly, we now fix the system parameters such that the phase diagram features a first-order phase transition at low $T$. We set $r=3.47$, $\mu=0.1$, $g=-1.4$. In this case a first-order quantum phase transition is located at $h_I =0.59106$ (at the mean-field level). The numerically obtained dependence of $\Gamma_q/\mu$ on $h$ is shown in Fig. \ref{Pic4} for $|\vec{q}|/\sqrt{2m\mu}=0.0179$. According to the results of Sec. III damping is expected to occur for $h\gtrsim\sqrt{\Delta^2+\mu^2}=0.563$, which very well agrees with the numerical data of Fig. \ref{Pic4}.
\section{Conclusion and outlook}

We have studied the damping process of the Goldstone mode for spin- and mass-imbalanced Fermi mixtures by inspecting the structure of the pair fluctuation propagator. A detailed analysis based on the gradient expansion  reveals presence of a temporally nonlocal contribution in its matrix elements, giving rise to Landau damping. We have demonstrated that the Landau damping is activated by increasing imbalance even for $T\to 0$ and is present for large enough mismatch of the Fermi surfaces. We have derived an analytical criterion for its occurrence  [see Eq.~(\ref{con2})]. We also provided an intuitive interpretation of the obtained analytical results. Finally, going beyond the gradient expansion, we have shown that our analytical predictions are in full agreement with damping rates obtained numerically from complex roots of the analytically continued determinant of the inverse pair fluctuation propagator.\\
\indent There are several interesting avenues for further research in this direction. The present analysis is performed at the Gaussian level (equivalent to the random phase approximation), under the assumption of the presence of fully-developed long-ranged order. It might be very interesting to investigate the evolution of the obtained physical picture after accounting for fluctuation effects. These should be of substantial relevance in particular in low dimensions, where the long-ranged ordered state becomes downgraded to the algebraic (Kosterlitz-Thouless) phase. Another question concerns the influence of the competing FFLO phase (characterized by nonzero ordering wavevector) on the excitation spectra. Even though theoretical results\cite{shimahara_1998, radzihovsky_2011, jakubczyk_2017, yin_2014, wang_2018} suggest that, in case of the neutral Fermi superfluids, these pair density wave states are unstable at $T>0$, they are presumably still present as ground states. We finally note the interesting question concerning the impact of imbalance on damping of the amplitude mode, whose existence was recently experimentally established.\cite{hoinka_2017, behrle_2018, liu_2016, salasnich_2017, castin_2019}
\begin{acknowledgments}
We acknowledge support from the Polish National Science Center via 
2014/15/B/ST3/02212 and 2017/26/E/ST3/00211. We thank Nicolas Dupuis for a~useful discussion. 
\end{acknowledgments}

\appendix

\section{Matrix elements of $\mathbb{F}^{-1}_q$}
In this appendix, we present the explicit form of the matrix elements of $\mathbb{F}_q^{-1}$ [see Eq. (\ref{M11}) and (\ref{M12})]. We assume without loss of generality, that $\Delta, u_k$ and $v_k \in\mathbb{R}$. The considered matrix elements are given by
\begin{align}
M_{1,1}(iq_m; \vec{q})&=\frac{1}{g}+\int_\vec{k}\sum_{\sigma, \sigma'}\mathcal{C}_{k, q}^{\sigma, \sigma'}\frac{f_{k, q}^{\sigma, \sigma'}}{iq_m-E_{k, q}^{\sigma, \sigma'}}, 
\label{M11A}\\
 M_{1,2}(iq_m; \vec{q})&=\int_\vec{k}\sum_{\sigma, \sigma'}\mathcal{D}_{k, q}^{\sigma, \sigma'}\frac{f_{k, q}^{\sigma, \sigma'}}{iq_m-E_{k, q}^{\sigma, \sigma'}},
\label{M12A}
\end{align}
where $E_{k, q}^{\sigma, \sigma'}=-(E_{-k}^\sigma-E^{\sigma'}_{k+q})$, $f_{k, q}^{\sigma, \sigma'}=f(E_{-k}^\sigma)-f(E_{k+q}^{\sigma'})$,  $\mathcal{D}_{k, q}^{\sigma, \sigma'}=\sigma\cdot\sigma'\cdot u_{-k}v_{-k}u_{k+q}v_{k+q}$ and
\begin{equation}
\mathcal{C}_{k, q}^{\sigma, \sigma'}=\begin{cases}
u_{k+q}^2v^2_{-k}; \;\;\;\;\; \sigma=+,\, \sigma'=+ , \\
u_{k+q}^2u^2_{-k}; \;\;\;\;\; \sigma=-,\, \sigma'=+ ,\\
v_{k+q}^2v^2_{-k}; \;\;\;\;\; \sigma=+,\, \sigma'=-,\\
v_{k+q}^2u^2_{-k}; \;\;\;\;\; \sigma=-,\, \sigma'=-.
\end{cases} \nonumber
\end{equation}
In the above expressions, we performed summation over fermionic Matsubara frequencies using standard textbook techniques\cite{bruus_2016, altland_2010, abrikosov_2016}.

\section{Gradient expansion of $\mathbb{F}^{-1}_q$}
In this appendix, we present the coefficients of the gradient expansion, which appears in Eqs. (\ref{M11E}) and (\ref{M12E}). They are given by
\begin{align}
A=&\int_\vec{k}\frac{1}{24 E_k^3} \Bigg[-4 E_k^3 u^2_k \Big(3 f''(E_k^-) \{a_k \alpha _k^--\delta _k^- v^2_k\}+\nonumber\\&
 6 b_k f'(E_k^-)-(\alpha _k^-)^2 f^{(3)}(E_k^-) v^2_k\Big)+\nonumber\\&
4 E_k^3 v^2_k \Big(3 f''(E_k^+) \{a_k \alpha_k^++\delta_k^+ u^2_k\}+\nonumber\\&
6 b_k f'(E_k^+)+(\alpha_k^+)^2 f^{(3)}(E_k^+) u^2_k\Big)+\nonumber\\&
3 u^2_k \Big([f(E_k^+)-f(E_k^-)] \{-2\alpha_k^+  a_k E_k+\nonumber\\&
4 b_k E_k^2-u^2_k\big(2\delta_k^+ E_k -(\alpha _k^+)^2\big)\}+\nonumber\\&
2\alpha_k^+ f'(E_k^+) E_k \{2 a_k E_k-\alpha_k^+ u^2_k\}+\nonumber\\&
2 E_k^2u^2_k \{\alpha _+^2 f''(E_k^+)+2\delta_k^+  f'(E_k^+)\}\Big)+\nonumber\\&
3 v^2_k \Big([f(E_k^-)-f(E_k^+)] \{2\alpha _k^-  a_k E_k+4 b_k E_k^2-\nonumber\\&
v^2_k \big((\alpha _k^-)^2+2\delta_k^- E_k\big)\}+\nonumber\\&
2\alpha_k^- f'(E_k^-) E_k \{2 a_k E_k-\alpha_k^- v^2_k\}-\nonumber\\&
2 E_k^2 v^2_k \{(\alpha_k^-)^2 f''(E_k^-)+2\delta_k^- f'(E_k^-)\}\Big)\Bigg]\nonumber,
\end{align}
$$B=\int_\vec{k}[f(E_k^+)-f(E_k^-)] \frac{u_k^4-v_k^4}{4 E_k^2},$$
$$C=\int_\vec{k}[f(E_k^-)-f(E_k^+)] \frac{ u_k^4+v_k^4}{8 E_k^3},$$
\begin{align}
D=&\int_\vec{k} \frac{u_kv_k }{6}\Big[-3\alpha_k^- d_k f''(E_k^-)-3\alpha_k^+  d_k f''(E_k^+)-\nonumber\\
&\frac{3}{4 E_k^3}\Big(2\alpha_k^- f'(E_k^-)  E_k \{2 d_k E_k-\alpha_k^- u_kv_k\}+\nonumber\\
&[f(E_k^-)-f(E_k^+)] \{2\alpha_k^- d_k E_k+4 E_k^2 g_k-\nonumber\\
&u_kv_k \{(\alpha_k^-)^2+2\delta_k^- E_k\}\}-\nonumber\\
&2 E_k^2 u_kv_k \{(\alpha_k^-)^2 f''(E_k^-)+2\delta_k^- f'(E_k^-)\}\Big)+\nonumber\\
&\frac{3}{4 E_k^3}\Big( 2\alpha_k^+ f'(E_k^+)  E_k \{2 d_k E_k+\alpha_k^+ u_kv_k\}+\nonumber\\&
[f(E_k^-)-f(E_k^+)] \{2\alpha_k^+ d_kE_k-4E_k^2 g_k-\nonumber\\
&u_kv_k\{2E_k\delta_k^+-(\alpha_k^+)^2\}\}-\nonumber\\
&2 E_k^2 u_kv_k \{(\alpha_k^+)^2 f''(E_k^+)+2\delta_k^+  f'(E_k^+)\}\Big)+\nonumber\\
&3 u_kv_k\{\delta_k^- f''(E_k^-)+\delta_k^+ f''(E_k^+)\}+\nonumber\\
&u_kv_k\{(\alpha_k^-)^2 f^{(3)}(E_k^-)+(\alpha_k^+)^2 f^{(3)}(E_k^+)\}-\nonumber\\
&6g_k\{f'(E_k^-)-f'(E_k^+)\}\Big]\nonumber
\end{align}
and
$$E=\int_\vec{k}[f(E_k^+)-f(E_k^-)] \frac{ u_k^2v_k^2}{4 E_k^3}.$$ In the above equations we use the following abbreviations: 
$\alpha_k=\xi_k|\vec{k}|\cos\theta/m E_k,$
$\delta_k=\xi_k\Delta^2/2mE_k^3+\xi^3_k/2mE_k^3+\Delta^2 |\vec{k}|^2\cos^2\theta/2mE_k^3,$
$\alpha^\sigma_k=\zeta|\vec{k}|\cos\theta/m+\sigma\alpha_k,$ $\delta^\sigma_k=\zeta/2m+\sigma\delta_k,$
$a_k=\Delta^2|\vec{k}|\cos\theta/2mE_k^3,$
$b_k=\Delta^4/4mE_k^5+\Delta^2\xi_k^2/4mE_k^5-3\Delta^2\xi_k|\vec{k}|^2\cos^2\theta/4m^2E_k^5,$
$d_k=\Delta\xi_k|\vec{k}|\cos\theta/2mE_k^3$ and
$g_k=\Delta(m\Delta^2\xi_k+m\xi_k^3+\Delta^2|\vec{k}|^2\cos^2\theta-2|\vec{k}|^2\xi_k^2\cos^2\theta)/4m^2E_k^5.$

%


\begin{thebibliography}{70}%
\makeatletter
\providecommand \@ifxundefined [1]{%
 \@ifx{#1\undefined}
}%
\providecommand \@ifnum [1]{%
 \ifnum #1\expandafter \@firstoftwo
 \else \expandafter \@secondoftwo
 \fi
}%
\providecommand \@ifx [1]{%
 \ifx #1\expandafter \@firstoftwo
 \else \expandafter \@secondoftwo
 \fi
}%
\providecommand \natexlab [1]{#1}%
\providecommand \enquote  [1]{``#1''}%
\providecommand \bibnamefont  [1]{#1}%
\providecommand \bibfnamefont [1]{#1}%
\providecommand \citenamefont [1]{#1}%
\providecommand \href@noop [0]{\@secondoftwo}%
\providecommand \href [0]{\begingroup \@sanitize@url \@href}%
\providecommand \@href[1]{\@@startlink{#1}\@@href}%
\providecommand \@@href[1]{\endgroup#1\@@endlink}%
\providecommand \@sanitize@url [0]{\catcode `\\12\catcode `\$12\catcode
  `\&12\catcode `\#12\catcode `\^12\catcode `\_12\catcode `\%12\relax}%
\providecommand \@@startlink[1]{}%
\providecommand \@@endlink[0]{}%
\providecommand \url  [0]{\begingroup\@sanitize@url \@url }%
\providecommand \@url [1]{\endgroup\@href {#1}{\urlprefix }}%
\providecommand \urlprefix  [0]{URL }%
\providecommand \Eprint [0]{\href }%
\providecommand \doibase [0]{http://dx.doi.org/}%
\providecommand \selectlanguage [0]{\@gobble}%
\providecommand \bibinfo  [0]{\@secondoftwo}%
\providecommand \bibfield  [0]{\@secondoftwo}%
\providecommand \translation [1]{[#1]}%
\providecommand \BibitemOpen [0]{}%
\providecommand \bibitemStop [0]{}%
\providecommand \bibitemNoStop [0]{.\EOS\space}%
\providecommand \EOS [0]{\spacefactor3000\relax}%
\providecommand \BibitemShut  [1]{\csname bibitem#1\endcsname}%
\let\auto@bib@innerbib\@empty
\bibitem [{\citenamefont {Bloch}\ \emph {et~al.}(2008)\citenamefont {Bloch},
  \citenamefont {Dalibard},\ and\ \citenamefont
  {Zwerger}}]{bloch_2008}
  \BibitemOpen
  \bibfield  {author} {\bibinfo {author} {\bibfnamefont {I.}~\bibnamefont
  {Bloch}}, \bibinfo {author} {\bibfnamefont {J.}~\bibnamefont {Dalibard}}, \
  and\ \bibinfo {author} {\bibfnamefont {W.}~\bibnamefont {Zwerger}},\ }\href
  {\doibase 10.1103/RevModPhys.80.885} {\bibfield  {journal} {\bibinfo
  {journal} {Rev. Mod. Phys.}\ }\textbf {\bibinfo {volume} {80}},\ \bibinfo
  {pages} {885} (\bibinfo {year} {2008})}\BibitemShut {NoStop}%
\bibitem [{\citenamefont {Giorgini}\ \emph {et~al.}(2008)\citenamefont {Giorgini},
  \citenamefont {Pitaevskii},\ and\ \citenamefont
  {Stringari}}]{giorgini_2008}
  \BibitemOpen
  \bibfield  {author} {\bibinfo {author} {\bibfnamefont {S.}~\bibnamefont
  {Giorgini}}, \bibinfo {author} {\bibfnamefont {L. P.}~\bibnamefont {Pitaevskii}}, \
  and\ \bibinfo {author} {\bibfnamefont {S.}~\bibnamefont {Stringari}},\ }\href
  {\doibase 10.1103/RevModPhys.80.1215} {\bibfield  {journal} {\bibinfo
  {journal} {Rev. Mod. Phys.}\ }\textbf {\bibinfo {volume} {80}},\ \bibinfo
  {pages} {1215} (\bibinfo {year} {2008})}\BibitemShut {NoStop}%
\bibitem [{\citenamefont {Mukherjee}\ \emph {et~al.}(2017)\citenamefont
  {Mukherjee}, \citenamefont {Yan}, \citenamefont {Patel}, \citenamefont
  {Hadzibabic}, \citenamefont {Yefsah}, \citenamefont {Struck},\ and\
  \citenamefont {Zwierlein}}]{murkherjee_17}
  \BibitemOpen
  \bibfield  {author} {\bibinfo {author} {\bibfnamefont {B.}~\bibnamefont
  {Mukherjee}}, \bibinfo {author} {\bibfnamefont {Z.}~\bibnamefont {Yan}},
  \bibinfo {author} {\bibfnamefont {P.~B.}\ \bibnamefont {Patel}}, \bibinfo
  {author} {\bibfnamefont {Z.}~\bibnamefont {Hadzibabic}}, \bibinfo {author}
  {\bibfnamefont {T.}~\bibnamefont {Yefsah}}, \bibinfo {author} {\bibfnamefont
  {J.}~\bibnamefont {Struck}}, \ and\ \bibinfo {author} {\bibfnamefont {M.~W.}\
  \bibnamefont {Zwierlein}},\ }\href {\doibase 10.1103/PhysRevLett.118.123401}
  {\bibfield  {journal} {\bibinfo  {journal} {Phys. Rev. Lett.}\ }\textbf
  {\bibinfo {volume} {118}},\ \bibinfo {pages} {123401} (\bibinfo {year}
  {2017})}\BibitemShut {NoStop}%
\bibitem [{\citenamefont {Hueck}\ \emph {et~al.}(2018)\citenamefont {Hueck},
  \citenamefont {Luick}, \citenamefont {Sobirey}, \citenamefont {Siegl},
  \citenamefont {Lompe},\ and\ \citenamefont
  {Moritz}}]{hueck_2018}
  \BibitemOpen
  \bibfield  {author} {\bibinfo {author} {\bibfnamefont {K.}~\bibnamefont
  {Hueck}}, \bibinfo {author} {\bibfnamefont {N.}~\bibnamefont {Luick}},
  \bibinfo {author} {\bibfnamefont {L.}~\bibnamefont {Sobirey}}, \bibinfo
  {author} {\bibfnamefont {J.}~\bibnamefont {Siegl}}, \bibinfo {author}
  {\bibfnamefont {T.}~\bibnamefont {Lompe}}, \ and\ \bibinfo {author}
  {\bibfnamefont {H.}~\bibnamefont {Moritz}},\ }\href {\doibase
  10.1103/PhysRevLett.120.060402} {\bibfield  {journal} {\bibinfo  {journal}
  {Phys. Rev. Lett.}\ }\textbf {\bibinfo {volume} {120}},\ \bibinfo {pages}
  {060402} (\bibinfo {year} {2018})}\BibitemShut {NoStop}%
\bibitem [{\citenamefont {Taglieber}\ \emph {et~al.}(2008)\citenamefont {Taglieber},
  \citenamefont {Voigt}, \citenamefont {Aoki}, \citenamefont {H\"ansch},\ and\ \citenamefont
  {Dieckmann}}]{taglieber_2008}
  \BibitemOpen
  \bibfield  {author} {\bibinfo {author} {\bibfnamefont {M.}~\bibnamefont
  {Taglieber}}, \bibinfo {author} {\bibfnamefont {A.~C.}~\bibnamefont {Voigt}}, \bibinfo
  {author} {\bibfnamefont {T.}~\bibnamefont {Aoki}},  \bibinfo
  {author} {\bibfnamefont {T.~W.}~\bibnamefont {H\"{a}nsch}}, \ and\ \bibinfo
  {author} {\bibfnamefont {K.}~\bibnamefont {Dieckmann}},\ }\href {\doibase
  10.1103/PhysRevLett.100.010401} {\bibfield  {journal} {\bibinfo  {journal}
  {Phys. Rev. Lett.}\ }\textbf {\bibinfo {volume} {100}},\ \bibinfo {pages}
  {010401} (\bibinfo {year} {2008})}\BibitemShut {NoStop}%
\bibitem [{\citenamefont {Wille}\ \emph {et~al.}(2008)\citenamefont {Wille},
  \citenamefont {Spiegelhalder}, \citenamefont {Kerner}, \citenamefont {Naik}, \citenamefont {Trenkwalder}, \citenamefont {Hendl}, \citenamefont {Schreck}, \citenamefont {Grimm}, \citenamefont {Tiecke}, \citenamefont {Walraven}, \citenamefont {Kokkelmans}, \citenamefont {Tiesinga},\ and\ \citenamefont
  {Julienne}}]{wille_2008}
  \BibitemOpen
  \bibfield  {author} {\bibinfo {author} {\bibfnamefont {E.}~\bibnamefont
  {Wille}}, \bibinfo {author} {\bibfnamefont {F.~M.}~\bibnamefont {Spiegelhalder}}, \bibinfo
  {author} {\bibfnamefont {G.}~\bibnamefont {Kerner}},  \bibinfo
  {author} {\bibfnamefont {D.}~\bibnamefont {Naik}}, \bibinfo
  {author} {\bibfnamefont {A.}~\bibnamefont {Trenkwalder}}, \bibinfo
  {author} {\bibfnamefont {G.}~\bibnamefont {Hendl}}, \bibinfo
  {author} {\bibfnamefont {F.}~\bibnamefont {Schreck}}, \bibinfo
  {author} {\bibfnamefont {R.}~\bibnamefont {Grimm}}, \bibinfo
  {author} {\bibfnamefont {T.~G.}~\bibnamefont {Tiecke}}, \bibinfo
  {author} {\bibfnamefont {J. T. M.} \bibnamefont {Walraven}}, \bibinfo
  {author} {\bibfnamefont {S. J. J. M. F.} \bibnamefont {Kokkelmans}}, \bibinfo
  {author} {\bibfnamefont {E.}~\bibnamefont {Tiesinga}}, \ and\ \bibinfo
  {author} {\bibfnamefont {P.~S.}~\bibnamefont {Julienne}},\ }\href {\doibase
  10.1103/PhysRevLett.100.053201} {\bibfield  {journal} {\bibinfo  {journal}
  {Phys. Rev. Lett.}\ }\textbf {\bibinfo {volume} {100}},\ \bibinfo {pages}
  {053201} (\bibinfo {year} {2008})}\BibitemShut {NoStop}%
\bibitem [{\citenamefont {Voigt}\ \emph {et~al.}(2009) \citenamefont {Voigt}, \citenamefont {Taglieber},  \citenamefont {Costa}, \citenamefont {Aoki}, \citenamefont {Wieser}, \citenamefont {H\"ansch},\ and\ \citenamefont
  {Dieckmann}}]{voigt_2009}
  \BibitemOpen
  \bibfield  {author} {\bibinfo {author} {\bibfnamefont {A. C.}~\bibnamefont
  {Voigt}}, \bibinfo {author} {\bibfnamefont {M.}~\bibnamefont {Taglieber}}, \bibinfo
  {author} {\bibfnamefont {L.}~\bibnamefont {Costa}}, \bibinfo
  {author} {\bibfnamefont {T.}~\bibnamefont {Aoki}}, \bibinfo
  {author} {\bibfnamefont {W.}~\bibnamefont {Wieser}}, \bibinfo
  {author} {\bibfnamefont {T.~W.}~\bibnamefont {H\"{a}nsch}}, \ and\ \bibinfo
  {author} {\bibfnamefont {K.}~\bibnamefont {Dieckmann}},\ }\href {\doibase
  10.1103/PhysRevLett.102.020405} {\bibfield  {journal} {\bibinfo  {journal}
  {Phys. Rev. Lett.}\ }\textbf {\bibinfo {volume} {102}},\ \bibinfo {pages}
  {020405} (\bibinfo {year} {2009})}\BibitemShut {NoStop}%
\bibitem [{\citenamefont {Tiecke}\ \emph {et~al.}(2008)\citenamefont {Tiecke},
  \citenamefont {Goosen}, \citenamefont {Ludewig}, \citenamefont {Gensemer}, \citenamefont {Kraft}, \citenamefont {Kokkelmans}, \ and\ \citenamefont {Walraven}}]{tiecke_2010}
  \BibitemOpen
  \bibfield  {author} {\bibinfo {author} {\bibfnamefont {T. G.}~\bibnamefont
  {Tiecke}}, \bibinfo {author} {\bibfnamefont {M. R.}~\bibnamefont {Goosen}}, \bibinfo
  {author} {\bibfnamefont {A.}~\bibnamefont {Ludewig}},  \bibinfo
  {author} {\bibfnamefont {S. D.}~\bibnamefont {Gensemer}}, \bibinfo
  {author} {\bibfnamefont {S.}~\bibnamefont {Kraft}}, \bibinfo
  {author} {\bibfnamefont {S. J. J. M. F.} \bibnamefont {Kokkelmans}}, \ and\ \bibinfo
  {author} {\bibfnamefont {J. T. M.} \bibnamefont {Walraven}},\ }\href {\doibase
  10.1103/PhysRevLett.104.053202} {\bibfield  {journal} {\bibinfo  {journal}
  {Phys. Rev. Lett.}\ }\textbf {\bibinfo {volume} {104}},\ \bibinfo {pages}
  {053202} (\bibinfo {year} {2010})}\BibitemShut {NoStop}%
\bibitem [{\citenamefont {Zwierlein}\ \emph {et~al.}(2006)\citenamefont
  {Zwierlein}, \citenamefont {Schirotzek}, \citenamefont {Schunck},\ and\
  \citenamefont {Ketterle}}]{zwierlein_2006}
  \BibitemOpen
  \bibfield  {author} {\bibinfo {author} {\bibfnamefont {M.~W.}\ \bibnamefont
  {Zwierlein}}, \bibinfo {author} {\bibfnamefont {A.}~\bibnamefont
  {Schirotzek}}, \bibinfo {author} {\bibfnamefont {C.~H.}\ \bibnamefont
  {Schunck}}, \ and\ \bibinfo {author} {\bibfnamefont {W.}~\bibnamefont
  {Ketterle}},\ }\href {\doibase 10.1126/science.1122318} {\bibfield  {journal}
  {\bibinfo  {journal} {Science}\ }\textbf {\bibinfo {volume} {311}},\ \bibinfo
  {pages} {492} (\bibinfo {year} {2006})}\BibitemShut {NoStop}%
\bibitem [{\citenamefont {Partridge}\ \emph {et~al.}(2006)\citenamefont
  {Partridge}, \citenamefont {Li}, \citenamefont {Kamar}, \citenamefont
  {Liao},\ and\ \citenamefont {Hulet}}]{partridge_2006}
  \BibitemOpen
  \bibfield  {author} {\bibinfo {author} {\bibfnamefont {G.~B.}\ \bibnamefont
  {Partridge}}, \bibinfo {author} {\bibfnamefont {W.}~\bibnamefont {Li}},
  \bibinfo {author} {\bibfnamefont {R.~I.}\ \bibnamefont {Kamar}}, \bibinfo
  {author} {\bibfnamefont {Y.-A.}\ \bibnamefont {Liao}}, \ and\ \bibinfo
  {author} {\bibfnamefont {R.~G.}\ \bibnamefont {Hulet}},\ }\href {\doibase
  10.1126/science.1122876} {\bibfield  {journal} {\bibinfo  {journal}
  {Science}\ }\textbf {\bibinfo {volume} {311}},\ \bibinfo {pages} {503}
  (\bibinfo {year} {2006})}\BibitemShut {NoStop}%
\bibitem [{\citenamefont {Ketterle}\ \emph {et~al.}(2009)\citenamefont
  {Ketterle}, \citenamefont {Shin}, \citenamefont {Schirotzek},\ and\
  \citenamefont {Schunk}}]{ketterle_2009}
  \BibitemOpen
  \bibfield  {author} {\bibinfo {author} {\bibfnamefont {W.}~\bibnamefont
  {Ketterle}}, \bibinfo {author} {\bibfnamefont {Y.}~\bibnamefont {Shin}},
  \bibinfo {author} {\bibfnamefont {A.}~\bibnamefont {Schirotzek}}, \ and\
  \bibinfo {author} {\bibfnamefont {C.~H.}\ \bibnamefont {Schunk}},\ }\href
  {http://stacks.iop.org/0953-8984/21/i=16/a=164206} {\bibfield  {journal}
  {\bibinfo  {journal} {J.~Phys.: Condens. Matter}\ }\textbf
  {\bibinfo {volume} {21}},\ \bibinfo {pages} {164206} (\bibinfo {year}
  {2009})}\BibitemShut {NoStop}%
\bibitem [{\citenamefont {Ong}\ \emph {et~al.}(2015)\citenamefont {Ong},
  \citenamefont {Cheng}, \citenamefont {Arakelyan},\ and\ \citenamefont
  {Thomas}}]{ong_2015}
  \BibitemOpen
  \bibfield  {author} {\bibinfo {author} {\bibfnamefont {W.}~\bibnamefont
  {Ong}}, \bibinfo {author} {\bibfnamefont {C.}~\bibnamefont {Cheng}}, \bibinfo
  {author} {\bibfnamefont {I.}~\bibnamefont {Arakelyan}}, \ and\ \bibinfo
  {author} {\bibfnamefont {J. E.}~\bibnamefont {Thomas}},\ }\href {\doibase
  10.1103/PhysRevLett.114.110403} {\bibfield  {journal} {\bibinfo  {journal}
  {Phys. Rev. Lett.}\ }\textbf {\bibinfo {volume} {114}},\ \bibinfo {pages}
  {110403} (\bibinfo {year} {2015})}\BibitemShut {NoStop}%
\bibitem [{\citenamefont {Mitra}\ \emph {et~al.}(2016)\citenamefont {Mitra},
  \citenamefont {Brown}, \citenamefont {Schauß}, \citenamefont {Kondov},\ and\
  \citenamefont {Bakr}}]{mitra_2016}
  \BibitemOpen
  \bibfield  {author} {\bibinfo {author} {\bibfnamefont {D.}~\bibnamefont
  {Mitra}}, \bibinfo {author} {\bibfnamefont {P.~T.}\ \bibnamefont {Brown}},
  \bibinfo {author} {\bibfnamefont {P.}~\bibnamefont {Schauss}}, \bibinfo
  {author} {\bibfnamefont {S.~S.}\ \bibnamefont {Kondov}}, \ and\ \bibinfo
  {author} {\bibfnamefont {W.~S.}\ \bibnamefont {Bakr}},\ }\href {\doibase
  10.1103/PhysRevLett.117.093601} {\bibfield  {journal} {\bibinfo  {journal}
  {Phys. Rev. Lett.}\ }\textbf {\bibinfo {volume} {117}},\ \bibinfo {pages}
  {093601} (\bibinfo {year} {2016})}\BibitemShut {NoStop}%
\bibitem [{\citenamefont {Sarma}(1963)}]{sarma_1963}
  \BibitemOpen
  \bibfield  {author} {\bibinfo {author} {\bibfnamefont {G.}~\bibnamefont
  {Sarma}},\ }\href {\doibase 10.1016/0022-3697(63)90007-6} {\bibfield
  {journal} {\bibinfo  {journal} {J. Phys. Chem. Solids}\ }\textbf {\bibinfo
  {volume} {24}},\ \bibinfo {pages} {1029} (\bibinfo {year}
  {1963})}\BibitemShut {NoStop}%
\bibitem [{\citenamefont {Liu}\ \emph {et~al.}(2003)\citenamefont
  {Liu},\ and\ \citenamefont {Wilczek}}]{liu_2003}
  \BibitemOpen
  \bibfield  {author} {\bibinfo {author} {\bibfnamefont {W.~V.}\ \bibnamefont
  {Liu}}  \ and\ \bibinfo {author} {\bibfnamefont {F.}~\bibnamefont
  {Wilczek}},\ }\href {\doibase 10.1103/PhysRevLett.90.047002} {\bibfield  {journal}
  {\bibinfo  {journal} {Phys. Rev. Lett.}\ }\textbf {\bibinfo {volume} {90}},\ \bibinfo
  {pages} {047002} (\bibinfo {year} {2003})}\BibitemShut {NoStop}%
\bibitem [{\citenamefont {Fulde}\ and\ \citenamefont
  {Ferrell}(1964)}]{fulde_1964}
  \BibitemOpen
  \bibfield  {author} {\bibinfo {author} {\bibfnamefont {P.}~\bibnamefont
  {Fulde}}\ and\ \bibinfo {author} {\bibfnamefont {R.~A.}\ \bibnamefont
  {Ferrell}},\ }\href {\doibase 10.1103/PhysRev.135.A550} {\bibfield  {journal}
  {\bibinfo  {journal} {Phys. Rev.}\ }\textbf {\bibinfo {volume} {135}},\
  \bibinfo {pages} {A550} (\bibinfo {year} {1964})} \BibitemShut {NoStop}%
\bibitem [{\citenamefont {Larkin}\ and\ \citenamefont
  {Ovchinnikov}(1965)}]{larkin_1965}
  \BibitemOpen
  \bibfield  {author} {\bibinfo {author} {\bibfnamefont {A.~I.}\ \bibnamefont
  {Larkin}}\ and\ \bibinfo {author} {\bibfnamefont {Y.~N.}\ \bibnamefont
  {Ovchinnikov}},\ }\href@noop {} {\bibfield  {journal} {\bibinfo  {journal}
  {Sov. Phys. JETP}\ }\textbf {\bibinfo {volume} {20}},\ \bibinfo {pages} {762}
  (\bibinfo {year} {1965})} \BibitemShut {NoStop}%
\bibitem [{\citenamefont {Parish}\ \emph
  {et~al.}(2007{\natexlab{b}})\citenamefont {Parish}, \citenamefont
  {Marchetti}, \citenamefont {Lamacraft},\ and\ \citenamefont
  {Simons}}]{parish_2007}
  \BibitemOpen
  \bibfield  {author} {\bibinfo {author} {\bibfnamefont {M.~M.}\ \bibnamefont
  {Parish}}, \bibinfo {author} {\bibfnamefont {F.~M.}\ \bibnamefont
  {Marchetti}}, \bibinfo {author} {\bibfnamefont {A.}~\bibnamefont
  {Lamacraft}}, \ and\ \bibinfo {author} {\bibfnamefont {B.~D.}\ \bibnamefont
  {Simons}},\ }\href {\doibase 10.1103/PhysRevLett.98.160402} {\bibfield
  {journal} {\bibinfo  {journal} {Phys. Rev. Lett.}\ }\textbf {\bibinfo
  {volume} {98}},\ \bibinfo {pages} {160402} (\bibinfo {year}
  {2007}{\natexlab{b}})}\BibitemShut {NoStop}%
\bibitem [{\citenamefont {Baarsma}\ \emph {et~al.}(2010)\citenamefont
  {Baarsma}, \citenamefont {Gubbels},\ and\ \citenamefont
  {Stoof}}]{baarsma_2010}
  \BibitemOpen
  \bibfield  {author} {\bibinfo {author} {\bibfnamefont {J.~E.}\ \bibnamefont
  {Baarsma}}, \bibinfo {author} {\bibfnamefont {K.~B.}\ \bibnamefont
  {Gubbels}}, \ and\ \bibinfo {author} {\bibfnamefont {H.~T.~C.}\ \bibnamefont
  {Stoof}},\ }\href {\doibase 10.1103/PhysRevA.82.013624} {\bibfield  {journal}
  {\bibinfo  {journal} {Phys. Rev. A}\ }\textbf {\bibinfo {volume} {82}},\
  \bibinfo {pages} {013624} (\bibinfo {year} {2010})}\BibitemShut {NoStop}%
\bibitem [{\citenamefont {Radzihovsky}\ and\ \citenamefont
  {Sheehy}(2010)}]{radzihovsky_2010}
  \BibitemOpen
  \bibfield  {author} {\bibinfo {author} {\bibfnamefont {L.}~\bibnamefont
  {Radzihovsky}}\ and\ \bibinfo {author} {\bibfnamefont {D.~E.}\ \bibnamefont
  {Sheehy}},\ }\href {\doibase 10.1088/0034-4885/73/7/076501} {\bibfield
  {journal} {\bibinfo  {journal} {Rep. Prog. Phys.}\ }\textbf {\bibinfo
  {volume} {73}},\ \bibinfo {pages} {076501} (\bibinfo {year}
  {2010})}\BibitemShut {NoStop}%
\bibitem [{\citenamefont {Strack}\ and\ \citenamefont
  {Jakubczyk}(2014)}]{strack_2014}
  \BibitemOpen
  \bibfield  {author} {\bibinfo {author} {\bibfnamefont {P.}~\bibnamefont
  {Strack}}\ and\ \bibinfo {author} {\bibfnamefont {P.}~\bibnamefont
  {Jakubczyk}},\ }\href {\doibase 10.1103/PhysRevX.4.021012} {\bibfield
  {journal} {\bibinfo  {journal} {Phys. Rev. X}\ }\textbf {\bibinfo {volume}
  {4}},\ \bibinfo {pages} {021012} (\bibinfo {year} {2014})}\BibitemShut
  {NoStop}%
\bibitem [{\citenamefont {Zdybel}\ \emph {et~al.}(2018)\citenamefont
  {Zdybel},\ and\ \citenamefont {Jakubczyk}}]{zdybel_2018}
  \BibitemOpen
  \bibfield  {author} {\bibinfo {author} {\bibfnamefont {P.}\ \bibnamefont
  {Zdybel}}  \ and\ \bibinfo {author} {\bibfnamefont {P.}~\bibnamefont
  {Jakubczyk}},\ }\href {\doibase 10.1088/1361-648X/aacc00} {\bibfield  {journal}
  {\bibinfo  {journal} {J. Phys.: Condens. Matter}\ }\textbf {\bibinfo {volume} {30}},\ \bibinfo
  {pages} {305604} (\bibinfo {year} {2018})} \BibitemShut {NoStop}%
\bibitem [{\citenamefont {Piazza}\ \emph {et~al.}(2016)\citenamefont {Piazza},
  \citenamefont {Zwerger},\ and\ \citenamefont {Strack}}]{piazza_2016}
  \BibitemOpen
  \bibfield  {author} {\bibinfo {author} {\bibfnamefont {F.}~\bibnamefont
  {Piazza}}, \bibinfo {author} {\bibfnamefont {W.}~\bibnamefont {Zwerger}}, \
  and\ \bibinfo {author} {\bibfnamefont {P.}~\bibnamefont {Strack}},\ }\href
  {\doibase 10.1103/PhysRevB.93.085112} {\bibfield  {journal} {\bibinfo
  {journal} {Phys. Rev. B}\ }\textbf {\bibinfo {volume} {93}},\ \bibinfo
  {pages} {085112} (\bibinfo {year} {2016})}\BibitemShut {NoStop}%
\bibitem [{\citenamefont {Pimenov}\ \emph {et~al.}(2017)\citenamefont
  {Pimenov}, \citenamefont {Mandal}, \citenamefont {Piazza},\ and\
  \citenamefont {Punk}}]{pimenov_2018}
  \BibitemOpen
  \bibfield  {author} {\bibinfo {author} {\bibfnamefont {D.}~\bibnamefont
  {Pimenov}}, \bibinfo {author} {\bibfnamefont {I.}~\bibnamefont {Mandal}},
  \bibinfo {author} {\bibfnamefont {F.}~\bibnamefont {Piazza}}, \ and\ \bibinfo
  {author} {\bibfnamefont {M.}~\bibnamefont {Punk}},\ }\href
  {\doibase 10.1103/PhysRevB.98.024510} {\bibfield  {journal} {\bibinfo
  {journal} {Phys. Rev. B}\ }\textbf {\bibinfo {volume} {98}},\ \bibinfo
  {pages} {024510} (\bibinfo {year} {2018})}\BibitemShut {NoStop}%
\bibitem [{\citenamefont {Iskin}\ \emph {et~al.}(2005)\citenamefont
  {Iskin},  \ and\ \citenamefont {S\'a de Melo}}]{iskin_2005}
  \BibitemOpen
  \bibfield  {author} {\bibinfo {author} {\bibfnamefont {M.}\ \bibnamefont
  {Iskin}} \ and\ \bibinfo {author} {\bibfnamefont {C. A. R.}~\bibnamefont
  {S\'a de Melo}},\ }\href {\doibase 10.1103/physrevb.72.024512} {\bibfield  {journal}
  {\bibinfo  {journal} {Phys. Rev. B}\ }\textbf {\bibinfo {volume} {72}},\ \bibinfo
  {pages} {024512} (\bibinfo {year} {2005})} \BibitemShut {NoStop}%
\bibitem [{\citenamefont {Iskin}\ \emph {et~al.}(2007)\citenamefont
  {Iskin},  \ and\ \citenamefont {S\'a de Melo}}]{iskin_2007a}
  \BibitemOpen
  \bibfield  {author} {\bibinfo {author} {\bibfnamefont {M.}\ \bibnamefont
  {Iskin}} \ and\ \bibinfo {author} {\bibfnamefont {C. A. R.}~\bibnamefont
  {S\'a de Melo}},\ }\href {\doibase 10.1007/s10909-007-9494-7} {\bibfield  {journal}
  {\bibinfo  {journal} {J. Low Temp. Phys.}\ }\textbf {\bibinfo {volume} {149}},\ \bibinfo
  {pages} {29} (\bibinfo {year} {2007})} \BibitemShut {NoStop}%
\bibitem [{\citenamefont {Klimin}\ \emph {et~al.}(2011)\citenamefont
  {Klimin}, \citenamefont
  {Tempere}, citenamefont
  {Lombardi}, \ and\ \citenamefont {Devreese}}]{klimin_2015}
  \BibitemOpen
  \bibfield  {author} {\bibinfo {author} {\bibfnamefont {S. N.}\ \bibnamefont
  {Klimin}}, \bibinfo {author} {\bibfnamefont {J.}\ \bibnamefont
  {Tempere}}, \bibinfo {author} {\bibfnamefont {G.}\ \bibnamefont
  {Lombardi}}, \ and\ \bibinfo {author} {\bibfnamefont {J. P. A.}~\bibnamefont
  {Devreese}},\ }\href {\doibase 10.1140/epjb/e2015-60213-4} {\bibfield  {journal}
  {\bibinfo  {journal} {Eur. Phys. J. B}\ }\textbf {\bibinfo {volume} {88}},\ \bibinfo
  {pages} {122} (\bibinfo {year} {2015})} \BibitemShut {NoStop}%
\bibitem [{\citenamefont {Klimin}\ \emph {et~al.}(2011)\citenamefont
  {Klimin}, \citenamefont
  {Tempere}, \ and\ \citenamefont {Kurkjian}}]{klimin_2019a}
  \BibitemOpen
  \bibfield  {author} {\bibinfo {author} {\bibfnamefont {S. N.}\ \bibnamefont
  {Klimin}}, \bibinfo {author} {\bibfnamefont {J.}\ \bibnamefont
  {Tempere}}, \ and\ \bibinfo {author} {\bibfnamefont {H.}~\bibnamefont
  {Kurkjian}},\ }\href {https://arxiv.org/abs/1908.11795v1} {\bibfield  {journal}
  {\bibinfo  {journal} {arXiv: 1908.11795v1}\ }(\bibinfo {year} {2019})} \BibitemShut {NoStop}%
\bibitem [{\citenamefont {Iskin}\ \emph {et~al.}(2011)\citenamefont
  {Iskin},  \ and\ \citenamefont {Suba\c{s}{\i}}}]{iskin_2011}
  \BibitemOpen
  \bibfield  {author} {\bibinfo {author} {\bibfnamefont {M.}\ \bibnamefont
  {Iskin}} \ and\ \bibinfo {author} {\bibfnamefont {A. L.}~\bibnamefont
  {Suba\c{s}{\i}}},\ }\href {\doibase 10.1103/PhysRevA.84.043621} {\bibfield  {journal}
  {\bibinfo  {journal} {Phys. Rev. A}\ }\textbf {\bibinfo {volume} {84}},\ \bibinfo
  {pages} {043621} (\bibinfo {year} {2011})} \BibitemShut {NoStop}%
\bibitem [{\citenamefont {Liao}\ \emph
  {et~al.}(2012{\natexlab{b}})\citenamefont {Liao}, \citenamefont
  {Yi-Xiang}, \ and\ \citenamefont
  {Liu}}]{liao_2012}
  \BibitemOpen
  \bibfield  {author} {\bibinfo {author} {\bibfnamefont {R.}\ \bibnamefont
  {Liao}}, \bibinfo {author} {\bibfnamefont {Y.}\ \bibnamefont
  {Yi-Xiang}},  \ and\ \bibinfo {author} {\bibfnamefont {W.-M.}\ \bibnamefont
  {Liu}},\ }\href {\doibase 10.1103/PhysRevLett.108.080406} {\bibfield
  {journal} {\bibinfo  {journal} {Phys. Rev. Lett.}\ }\textbf {\bibinfo
  {volume} {108}},\ \bibinfo {pages} {080406} (\bibinfo {year}
  {2012}{\natexlab{b}})}\BibitemShut {NoStop}%
\bibitem [{\citenamefont {Seo}\ \emph {et~al.}(2012)\citenamefont
  {Seo}, \citenamefont
  {Han},  \ and\ \citenamefont {S\'a de Melo}}]{seo_2012}
  \BibitemOpen
  \bibfield  {author} {\bibinfo {author} {\bibfnamefont {K.}\ \bibnamefont
  {Seo}}, \bibinfo {author} {\bibfnamefont {L.}\ \bibnamefont
  {Han}} \ and\ \bibinfo {author} {\bibfnamefont {C. A. R.}~\bibnamefont
  {S\'a de Melo}},\ }\href {\doibase 10.1103/PhysRevA.85.033601} {\bibfield  {journal}
  {\bibinfo  {journal} {Phys. Rev. A}\ }\textbf {\bibinfo {volume} {85}},\ \bibinfo
  {pages} {033601} (\bibinfo {year} {2012})} \BibitemShut {NoStop}%
\bibitem [{\citenamefont {Zhang}\ \emph
  {et~al.}(2013{\natexlab{b}})\citenamefont {Zhang}, \citenamefont
  {Yu}, \citenamefont
  {Ye},\ and\ \citenamefont
  {Liu}}]{zhang_2013}
  \BibitemOpen
  \bibfield  {author} {\bibinfo {author} {\bibfnamefont {S.-S.}\ \bibnamefont
  {Zhang}}, \bibinfo {author} {\bibfnamefont {X.-L.}\ \bibnamefont
  {Yu}}, \bibinfo {author} {\bibfnamefont {J.}\ \bibnamefont
  {Ye}}, \ and\ \bibinfo {author} {\bibfnamefont {W.-M.}\ \bibnamefont
  {Liu}},\ }\href {\doibase 10.1103/PhysRevA.87.063623} {\bibfield
  {journal} {\bibinfo  {journal} {Phys. Rev. A}\ }\textbf {\bibinfo
  {volume} {87}},\ \bibinfo {pages} {063623} (\bibinfo {year}
  {2013}{\natexlab{b}})}\BibitemShut {NoStop}%
\bibitem [{\citenamefont {Iskin}(1958)}]{iskin_2019}
  \BibitemOpen
  \bibfield  {author} {\bibinfo {author} {\bibfnamefont {M.}~\bibnamefont
  {Iskin}},\ }\href {https://arxiv.org/abs/1908.00818v1}
  {\bibfield  {journal} {\bibinfo  {journal} {arXiv:1908.00818v1}\ }(\bibinfo {year}
  {2019})}\BibitemShut {NoStop}%
\bibitem [{\citenamefont {Anderson}(1958)}]{anderson_1958}
  \BibitemOpen
  \bibfield  {author} {\bibinfo {author} {\bibfnamefont {P. W.}~\bibnamefont
  {Anderson}},\ }\href {\doibase 10.1103/PhysRev.112.1900}
  {\bibfield  {journal} {\bibinfo  {journal} {Phys. Rev.}\ }\textbf
  {\bibinfo {volume} {112}},\ \bibinfo {pages} {1900} (\bibinfo {year}
  {1958})}\BibitemShut {NoStop}%
\bibitem [{\citenamefont {Bogolybov}\ \emph {et~al.}(1958)\citenamefont
  {Bogolyubov}, \citenamefont
  {Tolmachev},\ and\ \citenamefont {Shirkov}}]{bogoliubov_1958}
  \BibitemOpen
  \bibfield  {author} {\bibinfo {author} {\bibfnamefont {N. N.}\ \bibnamefont
  {Bogolyubov}}, \bibinfo {author} {\bibfnamefont {V. V.}\ \bibnamefont
  {Tolmachev}},  \ and\ \bibinfo {author} {\bibfnamefont {D. V.}~\bibnamefont
  {Shirkov}},\ }\href {\doibase 10.1002/prop.19580061102} {\emph
  {\bibinfo {title} {A New Method in the Theory of Superconductivity}}}\ (\bibinfo
  {publisher} {Izd. Akad. Nauk SSSR [Engl. translation published by Consultants Bureau]},\ \bibinfo {year}
  {1958})\BibitemShut {NoStop}%
\bibitem [{\citenamefont {Nambu}(1960)}]{nambu_1960}
  \BibitemOpen
  \bibfield  {author} {\bibinfo {author} {\bibfnamefont {Y.}~\bibnamefont
  {Nambu}},\ }\href {\doibase 10.1103/PhysRev.117.648}
  {\bibfield  {journal} {\bibinfo  {journal} {Phys. Rev.}\ }\textbf
  {\bibinfo {volume} {117}},\ \bibinfo {pages} {648} (\bibinfo {year}
  {1960})}\BibitemShut {NoStop}%
\bibitem [{\citenamefont {Goldstone}(1961)}]{goldstone_1961}
  \BibitemOpen
  \bibfield  {author} {\bibinfo {author} {\bibfnamefont {J.}~\bibnamefont
  {Goldstone}},\ }\href {\doibase 10.1007/BF02812722}
  {\bibfield  {journal} {\bibinfo  {journal} {Il Nuovo Cimento}\ }\textbf
  {\bibinfo {volume} {19}},\ \bibinfo {pages} {154} (\bibinfo {year}
  {1961})}\BibitemShut {NoStop}%
\bibitem [{\citenamefont {Goldstone}\ \emph {et~al.}(1962)\citenamefont
  {Goldstone}, \citenamefont
  {Salam},\ and\ \citenamefont {Weinberg}}]{goldstone_1962}
  \BibitemOpen
  \bibfield  {author} {\bibinfo {author} {\bibfnamefont {J.}\ \bibnamefont
  {Goldstone}},  \bibinfo {author} {\bibfnamefont {A.}\ \bibnamefont
  {Salam}}, \ and\ \bibinfo {author} {\bibfnamefont {S.}~\bibnamefont
  {Weinberg}},\ }\href {\doibase 10.1103/PhysRev.127.965} {\bibfield  {journal}
  {\bibinfo  {journal} {Phys. Rev.}\ }\textbf {\bibinfo {volume} {127}},\ \bibinfo
  {pages} {965} (\bibinfo {year} {1962})} \BibitemShut {NoStop}%
\bibitem [{\citenamefont {Bartenstein}\ \emph {et~al.}(2004)\citenamefont
  {Bartenstein}, \citenamefont
  {Altmeyer}, \citenamefont
  {Riedl}, \citenamefont
  {Jochim}, \citenamefont
  {Chin}, \citenamefont
  {Denschlag}, \ and\ \citenamefont {Grimm}}]{bartenstein_2004}
  \BibitemOpen
  \bibfield  {author} {\bibinfo {author} {\bibfnamefont {M.}\ \bibnamefont
  {Bartenstein}},  \bibinfo {author} {\bibfnamefont {A.}\ \bibnamefont
  {Altmeyer}}, \bibinfo {author} {\bibfnamefont {S.}\ \bibnamefont
  {Riedl}}, \bibinfo {author} {\bibfnamefont {S.}\ \bibnamefont
  {Jochim}}, \bibinfo {author} {\bibfnamefont {C.}\ \bibnamefont
  {Chin}}, \bibinfo {author} {\bibfnamefont {J. H.}\ \bibnamefont
  {Denschlag}}, \ and\ \bibinfo {author} {\bibfnamefont {R.}~\bibnamefont
  {Grimm}},\ }\href {\doibase 10.1103/PhysRevLett.92.203201} {\bibfield  {journal}
  {\bibinfo  {journal} {Phys. Rev. Lett.}\ }\textbf {\bibinfo {volume} {92}},\ \bibinfo
  {pages} {203201} (\bibinfo {year} {2004})} \BibitemShut {NoStop}%
\bibitem [{\citenamefont {Altmeyer}\ \emph {et~al.}(2007) \citenamefont
  {Altmeyer}, \citenamefont
  {Riedl}, \citenamefont
  {Kohstall}, \citenamefont
  {Wright}, \citenamefont
  {Geursen}, \citenamefont
  {Bartenstein}, \citenamefont
  {Chin}, \citenamefont
  {Hecker Denschlag},\ and\ \citenamefont {Grimm}}]{altmeyer_2007}
  \BibitemOpen
  \bibfield  {author} {\bibinfo {author} {\bibfnamefont {A.}\ \bibnamefont
  {Altmeyer}}, \bibinfo {author} {\bibfnamefont {S.}\ \bibnamefont
  {Riedl}}, \bibinfo {author} {\bibfnamefont {C.}\ \bibnamefont
  {Kohstall}}, \bibinfo {author} {\bibfnamefont {M. J.}\ \bibnamefont
  {Wright}}, \bibinfo {author} {\bibfnamefont {R.}\ \bibnamefont
  {Geursen}}, \bibinfo {author} {\bibfnamefont {M.}\ \bibnamefont
  {Bartenstein}}, \bibinfo {author} {\bibfnamefont {C.}\ \bibnamefont
  {Chin}}, \bibinfo {author} {\bibfnamefont {J. H.}\ \bibnamefont
  {Denschlag}}, \ and\ \bibinfo {author} {\bibfnamefont {R.}~\bibnamefont
  {Grimm}},\ }\href {\doibase 10.1103/PhysRevLett.98.040401} {\bibfield  {journal}
  {\bibinfo  {journal} {Phys. Rev. Lett.}\ }\textbf {\bibinfo {volume} {98}},\ \bibinfo
  {pages} {040401} (\bibinfo {year} {2007})} \BibitemShut {NoStop}%
\bibitem [{\citenamefont {Tey}\ \emph {et~al.}(2013) \citenamefont
  {Tey}, \citenamefont
  {Sidorenkov}, \citenamefont
  {Guajardo}, \citenamefont
  {Grimm}, \citenamefont
  {Ku}, \citenamefont
  {Zwierlein}, \citenamefont
  {Hou}, \citenamefont
  {Pitaevskii},\ and\ \citenamefont {Stringari}}]{tey_2013}
  \BibitemOpen
  \bibfield  {author} {\bibinfo {author} {\bibfnamefont {M. K.}\ \bibnamefont
  {Tey}}, \bibinfo {author} {\bibfnamefont {L. A.}\ \bibnamefont
  {Sidorenkov}}, \bibinfo {author} {\bibfnamefont {Edmundo R. Sanchez}\ \bibnamefont
  {Guajardo}}, \bibinfo {author} {\bibfnamefont {R.}\ \bibnamefont
  {Grimm}}, \bibinfo {author} {\bibfnamefont {M. J. H.}\ \bibnamefont
  {Ku}}, \bibinfo {author} {\bibfnamefont {M. W.}\ \bibnamefont
  {Zwierlein}}, \bibinfo {author} {\bibfnamefont {Y.-H.}\ \bibnamefont
  {Hou}}, \bibinfo {author} {\bibfnamefont {L.}\ \bibnamefont
  {Pitaevskii}}, \ and\ \bibinfo {author} {\bibfnamefont {S.}~\bibnamefont
  {Stringari}},\ }\href {\doibase 10.1103/PhysRevLett.110.055303} {\bibfield  {journal}
  {\bibinfo  {journal} {Phys. Rev. Lett.}\ }\textbf {\bibinfo {volume} {110}},\ \bibinfo
  {pages} {055303} (\bibinfo {year} {2013})} \BibitemShut {NoStop}%
\bibitem [{\citenamefont {Sidorenkov}\ \emph {et~al.}(2013) \citenamefont
  {Sidorenkov}, \citenamefont
  {Tey}, \citenamefont
  {Grimm}, \citenamefont
  {Hou}, \citenamefont
  {Pitaevskii},\ and\ \citenamefont {Stringari}}]{sidorenkov_2013}
  \BibitemOpen
  \bibfield  {author} {\bibinfo {author} {\bibfnamefont {L. A.}\ \bibnamefont
  {Sidorenkov}}, \bibinfo {author} {\bibfnamefont {M. K.}\ \bibnamefont
  {Tey}}, \bibinfo {author} {\bibfnamefont {R.}\ \bibnamefont
  {Grimm}}, \bibinfo {author} {\bibfnamefont {Y.-H.}\ \bibnamefont
  {Hou}}, \bibinfo {author} {\bibfnamefont {L.}\ \bibnamefont
  {Pitaevskii}}, \ and\ \bibinfo {author} {\bibfnamefont {S.}~\bibnamefont
  {Stringari}},\ }\href {\doibase 10.1038/nature12136 } {\bibfield  {journal}
  {\bibinfo  {journal} {Nature}\ }\textbf {\bibinfo {volume} {498}},\ \bibinfo
  {pages} {78} (\bibinfo {year} {2013})} \BibitemShut {NoStop}%
\bibitem [{\citenamefont {Hoinka}\ \emph {et~al.}(2017) \citenamefont
  {Hoinka}, \citenamefont
  {Dyke}, \citenamefont
  {Lingham}, \citenamefont
  {Kinnunen}, \citenamefont
  {Bruun},\ and\ \citenamefont {Vale}}]{hoinka_2017}
  \BibitemOpen
  \bibfield  {author} {\bibinfo {author} {\bibfnamefont {S.}\ \bibnamefont
  {Hoinka}}, \bibinfo {author} {\bibfnamefont {P.}\ \bibnamefont
  {Dyke}}, \bibinfo {author} {\bibfnamefont {M. G.}\ \bibnamefont
  {Lingham}}, \bibinfo {author} {\bibfnamefont {J. J.}\ \bibnamefont
  {Kinnunen}}, \bibinfo {author} {\bibfnamefont {G. M.}\ \bibnamefont
  {Bruun}}, \ and\ \bibinfo {author} {\bibfnamefont {C. J.}~\bibnamefont
  {Vale}},\ }\href {\doibase 10.1038/nature12136 } {\bibfield  {journal}
  {\bibinfo  {journal} {Nature Phys.}\ }\textbf {\bibinfo {volume} {13}},\ \bibinfo
  {pages} {943} (\bibinfo {year} {2017})} \BibitemShut {NoStop}%
\bibitem [{\citenamefont {Engelbrecht}\ \emph {et~al.}(1997)\citenamefont
  {Engelbrecht}, \citenamefont
  {Randeria}, \ and\ \citenamefont {S\'{a} de Melo}}]{engelbrecht_1997}
  \BibitemOpen
  \bibfield  {author} {\bibinfo {author} {\bibfnamefont {J. R.}\ \bibnamefont
  {Engelbrecht}}, \bibinfo {author} {\bibfnamefont {M.}\ \bibnamefont
  {Randeria}}, \ and\ \bibinfo {author} {\bibfnamefont {C. A. R.}~\bibnamefont
  {S\'{a} de Melo}},\ }\href {\doibase 10.1103/physrevb.55.15153 } {\bibfield  {journal}
  {\bibinfo  {journal} {Phys. Rev. B}\ }\textbf {\bibinfo {volume} {55}},\ \bibinfo
  {pages} {15153} (\bibinfo {year} {1997})} \BibitemShut {NoStop}%
\bibitem [{\citenamefont {Marini}\ \emph {et~al.}(1998)\citenamefont
  {Marini}, \citenamefont
  {Pistolesi}, \ and\ \citenamefont {S\'{a} de Melo}}]{marini_1998}
  \BibitemOpen
  \bibfield  {author} {\bibinfo {author} {\bibfnamefont {M.}\ \bibnamefont
  {Marini}}, \bibinfo {author} {\bibfnamefont {F.}\ \bibnamefont
  {Pistolesi}}, \ and\ \bibinfo {author} {\bibfnamefont {G. C.}~\bibnamefont
  {Strinati}},\ }\href {\doibase 10.1007/s100510050165} {\bibfield  {journal}
  {\bibinfo  {journal} {Eur. Phys. J. B}\ }\textbf {\bibinfo {volume} {1}},\ \bibinfo
  {pages} {151} (\bibinfo {year} {1998})} \BibitemShut {NoStop}%
\bibitem [{\citenamefont {Ohashi}\ \emph {et~al.}(2003)\citenamefont
  {Ohashi},  \ and\ \citenamefont {Griffin}}]{ohashi_2003}
  \BibitemOpen
  \bibfield  {author} {\bibinfo {author} {\bibfnamefont {Y.}\ \bibnamefont
  {Ohashi}} \ and\ \bibinfo {author} {\bibfnamefont {A.}~\bibnamefont
  {Griffin}},\ }\href {\doibase 10.1103/physreva.67.063612} {\bibfield  {journal}
  {\bibinfo  {journal} {Phys. Rev. A}\ }\textbf {\bibinfo {volume} {67}},\ \bibinfo
  {pages} {063612} (\bibinfo {year} {2003})} \BibitemShut {NoStop}%
\bibitem [{\citenamefont {Combescot}\ \emph {et~al.}(2006)\citenamefont
  {Combescot}, \citenamefont
  {Kagan}, \ and\ \citenamefont {Stringari}}]{combescot_2006}
  \BibitemOpen
  \bibfield  {author} {\bibinfo {author} {\bibfnamefont {R.}\ \bibnamefont
  {Combescot}}, \bibinfo {author} {\bibfnamefont {M. Y.}\ \bibnamefont
  {Kagan}}, \ and\ \bibinfo {author} {\bibfnamefont {S.}~\bibnamefont
  {Stringari}},\ }\href {\doibase 10.1103/physreva.74.042717} {\bibfield  {journal}
  {\bibinfo  {journal} {Phys. Rev. A}\ }\textbf {\bibinfo {volume} {74}},\ \bibinfo
  {pages} {042717} (\bibinfo {year} {2006})} \BibitemShut {NoStop}%
\bibitem [{\citenamefont {Iskin}\ \emph {et~al.}(2007)\citenamefont
  {Iskin},  \ and\ \citenamefont {S\'a de Melo}}]{iskin_2007}
  \BibitemOpen
  \bibfield  {author} {\bibinfo {author} {\bibfnamefont {M.}\ \bibnamefont
  {Iskin}} \ and\ \bibinfo {author} {\bibfnamefont {C. A. R.}~\bibnamefont
  {S\'a de Melo}},\ }\href {\doibase 10.1103/physreva.76.013601} {\bibfield  {journal}
  {\bibinfo  {journal} {Phys. Rev. A}\ }\textbf {\bibinfo {volume} {76}},\ \bibinfo
  {pages} {013601} (\bibinfo {year} {2007})} \BibitemShut {NoStop}%
\bibitem [{\citenamefont {Hu}\ \emph {et~al.}(2007)\citenamefont
  {Hu}, \citenamefont
  {Drummond}, \ and\ \citenamefont {Liu}}]{hu_2007}
  \BibitemOpen
  \bibfield  {author} {\bibinfo {author} {\bibfnamefont {H.}\ \bibnamefont
  {Hu}}, \bibinfo {author} {\bibfnamefont {P. D.}\ \bibnamefont
  {Drummond}}, \ and\ \bibinfo {author} {\bibfnamefont {X.-J.}~\bibnamefont
  {Liu}},\ }\href {\doibase 10.1038/nphys598} {\bibfield  {journal}
  {\bibinfo  {journal} {Nature Phys.}\ }\textbf {\bibinfo {volume} {3}},\ \bibinfo
  {pages} {469} (\bibinfo {year} {2007})} \BibitemShut {NoStop}%
\bibitem [{\citenamefont {Diener}\ \emph {et~al.}(2008)\citenamefont
  {Diener}, \citenamefont
  {Sensarma}, \ and\ \citenamefont {Randeria}}]{diener_2008}
  \BibitemOpen
  \bibfield  {author} {\bibinfo {author} {\bibfnamefont {R. B.}\ \bibnamefont
  {Diener}}, \bibinfo {author} {\bibfnamefont {R.}\ \bibnamefont
  {Sensarma}}, \ and\ \bibinfo {author} {\bibfnamefont {M.}~\bibnamefont
  {Randeria}},\ }\href {\doibase 10.1103/physreva.77.023626} {\bibfield  {journal}
  {\bibinfo  {journal} {Phys. Rev. A}\ }\textbf {\bibinfo {volume} {77}},\ \bibinfo
  {pages} {023626} (\bibinfo {year} {2008})} \BibitemShut {NoStop}%
\bibitem [{\citenamefont {Klimin}\ \emph {et~al.}(2011)\citenamefont
  {Klimin}, \citenamefont
  {Tempere}, \ and\ \citenamefont {Devreese}}]{klimin_2011}
  \BibitemOpen
  \bibfield  {author} {\bibinfo {author} {\bibfnamefont {S. N.}\ \bibnamefont
  {Klimin}}, \bibinfo {author} {\bibfnamefont {J.}\ \bibnamefont
  {Tempere}}, \ and\ \bibinfo {author} {\bibfnamefont {J. P. A.}~\bibnamefont
  {Devreese}},\ }\href {\doibase 10.1007/s10909-011-0397-2} {\bibfield  {journal}
  {\bibinfo  {journal} {J. Low Temp. Phys.}\ }\textbf {\bibinfo {volume} {165}},\ \bibinfo
  {pages} {261} (\bibinfo {year} {2011})} \BibitemShut {NoStop}%
\bibitem [{\citenamefont {Kurkjian}\ \emph {et~al.}(2016)\citenamefont
  {Kurkjian}, \citenamefont
  {Castin}, \ and\ \citenamefont {Sinatra}}]{kurkjian_2016}
  \BibitemOpen
  \bibfield  {author} {\bibinfo {author} {\bibfnamefont {H.}\ \bibnamefont
  {Kurkjian}}, \bibinfo {author} {\bibfnamefont {Y.}\ \bibnamefont
  {Castin}}, \ and\ \bibinfo {author} {\bibfnamefont {A.}~\bibnamefont
  {Sinatra}},\ }\href {\doibase 10.1103/physreva.93.013623} {\bibfield  {journal}
  {\bibinfo  {journal} {Phys. Rev. A}\ }\textbf {\bibinfo {volume} {93}},\ \bibinfo
  {pages} {013623} (\bibinfo {year} {2016})} \BibitemShut {NoStop}%
\bibitem [{\citenamefont {Klimin}\ \emph {et~al.}(2011)\citenamefont
  {Klimin}, \citenamefont
  {Kurkjian}, \ and\ \citenamefont {Tempere}}]{tempere_2019}
  \BibitemOpen
  \bibfield  {author} {\bibinfo {author} {\bibfnamefont {S. N.}\ \bibnamefont
  {Klimin}}, \bibinfo {author} {\bibfnamefont {H.}\ \bibnamefont
  {Kurkjian}}, \ and\ \bibinfo {author} {\bibfnamefont {J.}~\bibnamefont
  {Tempere}},\ }\href {\doibase 10.1007/s10909-019-02160-3} {\bibfield  {journal}
  {\bibinfo  {journal} {J. Low Temp. Phys.}\ } (\bibinfo {year} {2019})} \BibitemShut {NoStop}%
\bibitem [{\citenamefont {Klimin}\ \emph {et~al.}(2011)\citenamefont
  {Klimin}, \citenamefont
  {Tempere}, \ and\ \citenamefont {Kurkjian}}]{klimin_2019}
  \BibitemOpen
  \bibfield  {author} {\bibinfo {author} {\bibfnamefont {S. N.}\ \bibnamefont
  {Klimin}}, \bibinfo {author} {\bibfnamefont {J.}\ \bibnamefont
  {Tempere}}, \ and\ \bibinfo {author} {\bibfnamefont {H.}~\bibnamefont
  {Kurkjian}},\ }\href {https://arxiv.org/abs/1811.07796v2} {\bibfield  {journal}
  {\bibinfo  {journal} {arXiv: 1811.07796v2}\ }(\bibinfo {year} {2019})} \BibitemShut {NoStop}%
\bibitem [{\citenamefont {Zou}\ \emph {et~al.}(2018)\citenamefont
  {Zou}, \citenamefont
  {Hu}, \ and\ \citenamefont {Liu}}]{zou_2018}
  \BibitemOpen
  \bibfield  {author} {\bibinfo {author} {\bibfnamefont {P.}\ \bibnamefont
  {Zou}}, \bibinfo {author} {\bibfnamefont {H.}\ \bibnamefont
  {Hu}}, \ and\ \bibinfo {author} {\bibfnamefont {X.-J.}~\bibnamefont
  {Liu}},\ }\href {\doibase 10.1103/PhysRevA.98.011602} {\bibfield  {journal}
  {\bibinfo  {journal} {Phys. Rev. A}\ }\textbf {\bibinfo {volume} {98}},\ \bibinfo
  {pages} {011602(R)} (\bibinfo {year} {2018})} \BibitemShut {NoStop}%
\bibitem [{\citenamefont {Shen}\ \emph {et~al.}(2015)\citenamefont
  {Shen}, \ and\ \citenamefont {Zheng}}]{shen_2015}
  \BibitemOpen
  \bibfield  {author} {\bibinfo {author} {\bibfnamefont {H.}\ \bibnamefont
  {Shen}}  \ and\ \bibinfo {author} {\bibfnamefont {W.}~\bibnamefont
  {Zheng}},\ }\href {\doibase 10.1103/PhysRevA.92.033620} {\bibfield  {journal}
  {\bibinfo  {journal} {Phys. Rev. A}\ }\textbf {\bibinfo {volume} {92}},\ \bibinfo
  {pages} {033620} (\bibinfo {year} {2015})} \BibitemShut {NoStop}%
\bibitem [{\citenamefont {Zhang}\ \emph {et~al.}(2011)\citenamefont
  {Zhang}, \ and\ \citenamefont {Liu}}]{zhang_2011}
  \BibitemOpen
  \bibfield  {author} {\bibinfo {author} {\bibfnamefont {Z.}\ \bibnamefont
  {Zhang}}  \ and\ \bibinfo {author} {\bibfnamefont {W. V.}~\bibnamefont
  {Liu}},\ }\href {\doibase 10.1103/PhysRevA.83.023617} {\bibfield  {journal}
  {\bibinfo  {journal} {Phys. Rev. A}\ }\textbf {\bibinfo {volume} {83}},\ \bibinfo
  {pages} {023617} (\bibinfo {year} {2011})} \BibitemShut {NoStop}%
\bibitem [{\citenamefont {Kurkjian}\ \emph {et~al.}(2017)\citenamefont
  {Kurkjian} \ and\ \citenamefont {Tempere}}]{kurkjian_2017}
  \BibitemOpen
  \bibfield  {author} {\bibinfo {author} {\bibfnamefont {H.}\ \bibnamefont
  {Kurkjian}}  \ and\ \bibinfo {author} {\bibfnamefont {J.}~\bibnamefont
  {Tempere}},\ }\href {\doibase 10.1088/1367-2630/aa969b} {\bibfield  {journal}
  {\bibinfo  {journal} {New J. Phys.}\ }\textbf {\bibinfo {volume} {19}},\ \bibinfo
  {pages} {113045} (\bibinfo {year} {2017})} \BibitemShut {NoStop}%
\bibitem [{\citenamefont {Bruus}(2016)}]{bruus_2016}
  \BibitemOpen
  \bibfield  {author} {\bibinfo {author} {\bibfnamefont {H.}~\bibnamefont
  {Bruus}}  \ and\ \bibinfo {author} {\bibfnamefont {K.}~\bibnamefont
  {Flensberg}},\ } {\emph
  {\bibinfo {title} {Many-Body Quantum Theory in Condensed Matter Physics: An Introduction}}}\ (\bibinfo
  {publisher} {Oxford University Press},\ \bibinfo {year}
  {2016})\BibitemShut {NoStop}%
\bibitem [{\citenamefont {Matera}\ \emph {et~al.}(2017)\citenamefont
  {Matera}, \ and\ \citenamefont {Wagner}}]{matera_2017}
  \BibitemOpen
  \bibfield  {author} {\bibinfo {author} {\bibfnamefont {F.}\ \bibnamefont
  {Matera}}  \ and\ \bibinfo {author} {\bibfnamefont {M. F.}~\bibnamefont
  {Wagner}},\ }\href {\doibase 10.1140/epjd/e2017-80156-0} {\bibfield  {journal}
  {\bibinfo  {journal} {Eur. Phys. J. D}\ }\textbf {\bibinfo {volume} {71}},\ \bibinfo
  {pages} {293} (\bibinfo {year} {2017})} \BibitemShut {NoStop}%
\bibitem [{\citenamefont {Altland}(2010)}]{altland_2010}
  \BibitemOpen
  \bibfield  {author} {\bibinfo {author} {\bibfnamefont {A.}~\bibnamefont
  {Altland}}  \ and\ \bibinfo {author} {\bibfnamefont {B.}~\bibnamefont
  {Simons}},\ } {\emph
  {\bibinfo {title} {Condensed Matter Field Theory}}}\ (\bibinfo
  {publisher} {Cambridge University Press},\ \bibinfo {year}
  {2010})\BibitemShut {NoStop}%
\bibitem [{\citenamefont {Nozi\'eres}\ \emph {et~al.}(1985)\citenamefont
  {Nozi\'eres}, \ and\ \citenamefont {Schmitt-Rink}}]{nozieres_1985}
  \BibitemOpen
  \bibfield  {author} {\bibinfo {author} {\bibfnamefont {P.}\ \bibnamefont
  {Nozi\'eres}} \ and\ \bibinfo {author} {\bibfnamefont {S.}~\bibnamefont
  {Schmitt-Rink}},\ }\href {\doibase 10.1007/bf00683774} {\bibfield  {journal}
  {\bibinfo  {journal} {J. Low Temp. Phys.}\ }\textbf {\bibinfo {volume} {59}},\ \bibinfo
  {pages} {195} (\bibinfo {year} {1985})} \BibitemShut {NoStop}%
\bibitem [{\citenamefont {Pieri}\ \emph {et~al.}(2004)\citenamefont
  {Pieri}, \citenamefont
  {Pisani}, \ and\ \citenamefont {Strinati}}]{pieri_2004}
  \BibitemOpen
  \bibfield  {author} {\bibinfo {author} {\bibfnamefont {P.}\ \bibnamefont
  {Pieri}}, \bibinfo {author} {\bibfnamefont {L.}\ \bibnamefont
  {Pisani}}, \ and\ \bibinfo {author} {\bibfnamefont {G. C.}~\bibnamefont
  {Strinati}},\ }\href {\doibase 10.1103/physrevb.70.094508} {\bibfield  {journal}
  {\bibinfo  {journal} {Phys. Rev. B}\ }\textbf {\bibinfo {volume} {70}},\ \bibinfo
  {pages} {094508} (\bibinfo {year} {2004})} \BibitemShut {NoStop}%
\bibitem [{\citenamefont {Abrikosov}(2016)}]{abrikosov_2016}
  \BibitemOpen
  \bibfield  {author} {\bibinfo {author} {\bibfnamefont {A. A.}~\bibnamefont
  {Abrikosov}}, \bibinfo {author} {\bibfnamefont {L. P.}~\bibnamefont
  {Gorkov}},  \ and\ \bibinfo {author} {\bibfnamefont {I. E.}~\bibnamefont
  {Dzyaloshinski}},\ } {\emph
  {\bibinfo {title} {Methods of Quantum Field Theory in Statistical Physics}}}\ (\bibinfo
  {publisher} {Dover Publ.},\ \bibinfo {year}
  {2016})\BibitemShut {NoStop}%
\bibitem [{\citenamefont {Hertz} (1976)}]{hertz_1976}
  \BibitemOpen
  \bibfield  {author} {\bibinfo {author} {\bibfnamefont {J. A.}\ \bibnamefont
  {Hertz}} \ }\href {\doibase 10.1103/PhysRevB.14.1165} {\bibfield  {journal}
  {\bibinfo  {journal} {Phys. Rev. B}\ }\textbf {\bibinfo {volume} {14}},\ \bibinfo
  {pages} {1165} (\bibinfo {year} {1976})} \BibitemShut {NoStop}%
\bibitem [{\citenamefont {Nagaosa}(1998)}]{nagaosa_1998}
  \BibitemOpen
  \bibfield  {author} {\bibinfo {author} {\bibfnamefont {N.}~\bibnamefont
  {Nagaosa}},\ } {\emph
  {\bibinfo {title} {Quantum Field Theory in Strongly Correlated Electronic Systems}}}\ (\bibinfo
  {publisher} {Springer},\ \bibinfo {year}
  {1998})\BibitemShut {NoStop}%
\bibitem [{\citenamefont {L\"ohneysen}\ \emph {et~al.}(2007)\citenamefont {L\"ohneysen},
  \citenamefont {Rosch},   \citenamefont {Vojta},\ and\ \citenamefont
  {W\"olfle}}]{lohneysen_2007}
  \BibitemOpen
  \bibfield  {author} {\bibinfo {author} {\bibfnamefont {H.}~\bibnamefont
  {v. L\"ohneysen}}, \bibinfo {author} {\bibfnamefont {A.}~\bibnamefont {Rosch}}, \bibinfo {author} {\bibfnamefont {M.}~\bibnamefont {Vojta}}, \ and\ \bibinfo {author} {\bibfnamefont {P.}~\bibnamefont {W\"olfle}},\ }\href
  {\doibase 10.1103/revmodphys.79.1015} {\bibfield  {journal} {\bibinfo
  {journal} {Rev. Mod. Phys.}\ }\textbf {\bibinfo {volume} {79}},\ \bibinfo
  {pages} {1015} (\bibinfo {year} {2007})}\BibitemShut {NoStop}%
\bibitem [{\citenamefont {Continentino}(2017)}]{continentino_2017}
  \BibitemOpen
  \bibfield  {author} {\bibinfo {author} {\bibfnamefont {M.}~\bibnamefont
  {Continentino}},\ } {\emph
  {\bibinfo {title} {Quantum Scaling in Many-Body Systems. An Approach to Quantum Phase Transitions}}}\ (\bibinfo
  {publisher} {Cambridge University Press},\ \bibinfo {year}
  {2017})\BibitemShut {NoStop}%
\bibitem [{\citenamefont {Nozi\`eres}(2018)}]{nozieres_2018}
  \BibitemOpen
  \bibfield  {author} {\bibinfo {author} {\bibfnamefont {P.}~\bibnamefont
  {Nozi\`eres}},\ } {\emph
  {\bibinfo {title} {Theory of Interacting Fermi Systems}}}\ (\bibinfo
  {publisher} {CRC Press},\ \bibinfo {year}
  {2018})\BibitemShut {NoStop}%
\bibitem [{\citenamefont {Shimahara}(1998)}]{shimahara_1998}
  \BibitemOpen
  \bibfield  {author} {\bibinfo {author} {\bibfnamefont {H.}~\bibnamefont
  {Shimahara}},\ }\href {\doibase 10.1143/JPSJ.67.1872}
  {\bibfield  {journal} {\bibinfo  {journal} {J. Phys. Soc. Jpn}\ }\textbf
  {\bibinfo {volume} {67}},\ \bibinfo {pages} {1872} (\bibinfo {year}
  {1998})}\BibitemShut {NoStop}%
\bibitem [{\citenamefont {Radzihovsky}(2011)}]{radzihovsky_2011}
  \BibitemOpen
  \bibfield  {author} {\bibinfo {author} {\bibfnamefont {L.}~\bibnamefont
  {Radzihovsky}},\ }\href {\doibase 10.1103/PhysRevA.84.023611}
  {\bibfield  {journal} {\bibinfo  {journal} {Phys. Rev. A}\ }\textbf
  {\bibinfo {volume} {84}},\ \bibinfo {pages} {023611} (\bibinfo {year}
  {2011})}\BibitemShut {NoStop}%
\bibitem [{\citenamefont {Yin}\ \emph
  {et~al.}(2014{\natexlab{b}})\citenamefont {Yin} , \citenamefont {Martikainen},\ and\ \citenamefont
  {T\"orm\"a}}]{yin_2014}
  \BibitemOpen
  \bibfield  {author} {\bibinfo {author} {\bibfnamefont {S.}\ \bibnamefont
  {Yin}},  \bibinfo {author} {\bibfnamefont {J.-P.}~\bibnamefont
  {Martikainen}}, \ and\ \bibinfo {author} {\bibfnamefont {P.}\ \bibnamefont
  {T\"orm\"a}},\ }\href {\doibase 10.1103/physrevb.89.014507} {\bibfield
  {journal} {\bibinfo  {journal} {Phys. Rev. B}\ }\textbf {\bibinfo
  {volume} {89}},\ \bibinfo {pages} {014507} (\bibinfo {year}
  {2014}{\natexlab{b}})}\BibitemShut {NoStop}%
\bibitem [{\citenamefont {Jakubczyk}(2017)}]{jakubczyk_2017}
  \BibitemOpen
  \bibfield  {author} {\bibinfo {author} {\bibfnamefont {P.}~\bibnamefont
  {Jakubczyk}},\ }\href {\doibase 10.1103/PhysRevA.95.063626}
  {\bibfield  {journal} {\bibinfo  {journal} {Phys. Rev. A}\ }\textbf
  {\bibinfo {volume} {95}},\ \bibinfo {pages} {063626} (\bibinfo {year}
  {2017})}\BibitemShut {NoStop}%
\bibitem [{\citenamefont {Wang}\ \emph
  {et~al.}(2014{\natexlab{b}})\citenamefont {Wang} , \citenamefont {Che}, \citenamefont {Zhang}, \ and\ \citenamefont
  {Chen}}]{wang_2018}
  \BibitemOpen
  \bibfield  {author} {\bibinfo {author} {\bibfnamefont {J.}\ \bibnamefont
  {Wang}},  \bibinfo {author} {\bibfnamefont {Y.}~\bibnamefont
  {Che}}, \bibinfo {author} {\bibfnamefont {L.}~\bibnamefont
  {Zhang}}, \ and\ \bibinfo {author} {\bibfnamefont {Q.}\ \bibnamefont
  {Chen}},\ }\href {\doibase 10.1103/physrevb.97.134513} {\bibfield
  {journal} {\bibinfo  {journal} {Phys. Rev. B}\ }\textbf {\bibinfo
  {volume} {97}},\ \bibinfo {pages} {134513} (\bibinfo {year}
  {2018}{\natexlab{b}})}\BibitemShut {NoStop}%
\bibitem [{\citenamefont {Behrle}\ \emph {et~al.}(2018) \citenamefont
  {Behrle}, \citenamefont
  {Harrison}, \citenamefont
  {Kombe}, \citenamefont
  {Gao}, \citenamefont
  {Link},\citenamefont
  {Bernier},\citenamefont
  {Kollath},\ and\ \citenamefont {K\"ohl}}]{behrle_2018}
  \BibitemOpen
  \bibfield  {author} {\bibinfo {author} {\bibfnamefont {A.}\ \bibnamefont
  {Behrle}}, \bibinfo {author} {\bibfnamefont {T.}\ \bibnamefont
  {Harrison}}, \bibinfo {author} {\bibfnamefont {J.}\ \bibnamefont
  {Kombe}}, \bibinfo {author} {\bibfnamefont {K.}\ \bibnamefont
  {Gao}}, \bibinfo {author} {\bibfnamefont {M.}\ \bibnamefont
  {Link}}, \bibinfo {author} {\bibfnamefont {J.-S}\ \bibnamefont
  {Bernier}}, \bibinfo {author} {\bibfnamefont {C.}\ \bibnamefont
  {Kollath}},\ and\ \bibinfo {author} {\bibfnamefont {M.}~\bibnamefont
  {K\"ohl}},\ }\href {\doibase 10.1038/s41567-018-0128-6} {\bibfield  {journal}
  {\bibinfo  {journal} {Nature Phys.}\ }\textbf {\bibinfo {volume} {14}},\ \bibinfo
  {pages} {781} (\bibinfo {year} {2018})} \BibitemShut {NoStop}%
\bibitem [{\citenamefont {Liu}\ \emph {et~al.}(2016)\citenamefont {Liu},
  \citenamefont {Zhai},\ and\ \citenamefont {Zhang}}]{liu_2016}
  \BibitemOpen
  \bibfield  {author} {\bibinfo {author} {\bibfnamefont {B.}~\bibnamefont
  {Liu}}, \bibinfo {author} {\bibfnamefont {H.}~\bibnamefont {Zhai}}, \ and\ \bibinfo {author} {\bibfnamefont {S.}~\bibnamefont {Zhang}},\ }\href
  {\doibase 10.1103/PhysRevA.93.033641} {\bibfield  {journal} {\bibinfo
  {journal} {Phys. Rev. A}\ }\textbf {\bibinfo {volume} {93}},\ \bibinfo
  {pages} {033641} (\bibinfo {year} {2016})}\BibitemShut {NoStop}%
\bibitem [{\citenamefont {Salasnich}(2017)}]{salasnich_2017}
  \BibitemOpen
  \bibfield  {author} {\bibinfo {author} {\bibfnamefont {L.}~\bibnamefont
  {Salasnich}},\ }\href {\doibase 10.3390/condmat2020022}
  {\bibfield  {journal} {\bibinfo  {journal} {Condens. Matter }\ }\textbf
  {\bibinfo {volume} {2}},\ \bibinfo {pages} {22} (\bibinfo {year}
  {2017})}\BibitemShut {NoStop}%
\bibitem [{\citenamefont {Kurkjian}\ \emph
  {et~al.}(2019{\natexlab{b}})\citenamefont {Kurkjian}, \citenamefont
  {Klimin}, \citenamefont {Tempere},\ and\ \citenamefont
  {Castin}}]{castin_2019}
  \BibitemOpen
  \bibfield  {author} {\bibinfo {author} {\bibfnamefont {H.}\ \bibnamefont
  {Kurkjian}}, \bibinfo {author} {\bibfnamefont {S.~N.}\ \bibnamefont
  {Klimin}}, \bibinfo {author} {\bibfnamefont {J.}~\bibnamefont
  {Tempere}}, \ and\ \bibinfo {author} {\bibfnamefont {Y.}\ \bibnamefont
  {Castin}},\ }\href {\doibase 10.1103/PhysRevLett.122.093403} {\bibfield
  {journal} {\bibinfo  {journal} {Phys. Rev. Lett.}\ }\textbf {\bibinfo
  {volume} {122}},\ \bibinfo {pages} {093403} (\bibinfo {year}
  {2019}{\natexlab{b}})}\BibitemShut {NoStop}%
\end{thebibliography}
\end{document}